\documentclass[12pt, draftclsnofoot,journal,onecolumn]{IEEEtran}
\usepackage{epsfig,epsf}
\usepackage{color}

\usepackage{bbold}
\usepackage{multirow}
\usepackage{latexsym}
\usepackage{amssymb}
\usepackage{amsmath}

\usepackage{amsfonts}
\usepackage[sort]{cite}
\usepackage{fullpage}
\usepackage{mathrsfs}
\usepackage{scalefnt}
\usepackage{extarrows}
\usepackage{setspace}
\usepackage{stfloats}
\usepackage{url}
\usepackage{hyphenat}
\usepackage{graphicx}
\usepackage[table]{xcolor}
\usepackage{bbm}
\usepackage{dsfont}
\usepackage{bm}
\usepackage{array}
\usepackage{wrapfig}
\usepackage[]{algpseudocode}
\usepackage[linesnumbered,ruled]{algorithm2e}
\newenvironment{Process}[1][htb]
  {
  \begin{algorithm}[#1]%
  }{\end{algorithm}}
\let\oldnl\nl
\newcommand{\nonl}{\renewcommand{\nl}{\let\nl\oldnl}}
\renewcommand{\algorithmicrequire}{\textbf{Input:}}
\renewcommand{\algorithmicensure}{\textbf{Output:}}
\usepackage{mathtools}
\allowdisplaybreaks[1]
\usepackage{epstopdf}
\newcommand\numberthis{\addtocounter{equation}{1}\tag{\theequation}}

\newsavebox\myboxA
\newsavebox\myboxB
\newlength\mylenA
\newcommand*\xoverline[2][0.75]{%
    \sbox{\myboxA}{$\m@th#2$}%
    \setbox\myboxB\null
    \ht\myboxB=\ht\myboxA%
    \dp\myboxB=\dp\myboxA%
    \wd\myboxB=#1\wd\myboxA
    \sbox\myboxB{$\m@th\overline{\copy\myboxB}$}
    \setlength\mylenA{\the\wd\myboxA}
    \addtolength\mylenA{-\the\wd\myboxB}%
    \ifdim\wd\myboxB<\wd\myboxA%
       \rlap{\hskip 0.5\mylenA\usebox\myboxB}{\usebox\myboxA}%
    \else
        \hskip -0.5\mylenA\rlap{\usebox\myboxA}{\hskip 0.5\mylenA\usebox\myboxB}%
    \fi}
\makeatother  
\usepackage{enumerate}

\newcommand\norm[1]{\left\lVert#1\right\rVert}

\def\phi{\varphi}

\def\He{{\mathsf{H}}}

\def\cC{{\cal C}}

\newcommand{\comm}[1]{}

\def\bff{{\mathbf{f}}}

\def\bv{{\mathbf{v}}}

\def\bx{{\mathbf{x}}}
\def\by{{\mathbf{y}}}
\def\bz{{\mathbf{z}}}
\def\b0{{\mathbf{0}}}

\def\argmax{\mathop{\mathrm{argmax}}}

\def\bQ{{\mathbf{Q}}}

\def\cC{\mathcal{C}}

\def\cE{\mathcal{E}}
\def\cF{\mathcal{F}}

\def\cH{\mathcal{H}}
\def\cI{\mathcal{I}}

\def\cN{\mathcal{N}}

\def\cP{\mathcal{P}}

\def\cR{\mathcal{R}}

\def\cX{\mathcal{X}}
\def\cY{\mathcal{Y}}

\def \Re[#1]{\text{Re}\left(#1\right)}
\def \Im[#1]{\text{Im}\left(#1\right)}
\def \Cpx[#1]{\tilde{#1}}
\def \tx {x} 
\def \rx {y} 

\def \txv {\bx} 
\def \rxv {\by} 

\def \txm {X} 
\def \rxm {Y} 

\def \txA {\cX} 
\def \rxA {\cY} 
\def \chm {G} 

\def \txbit {a}

\def \txvte[#1]{\txv_{\rm t,#1}} 
\def \rxvte[#1]{\rxv_{\rm t,#1}} 
\def \txvde[#1]{\txv_{\rm d,#1}} 
\def \rxvde[#1]{\rxv_{\rm d,#1}} 
\def \nvbe[#1]{\nv_{\rm b,#1}} 

\def \txde[#1]{\tx_{\rm d,#1}} 
\def \rxde[#1]{\rx_{\rm d,#1}} 

\def \txta[#1]{\tx_{{\rm t},#1}} 
\def \txda[#1]{\tx_{{\rm d},#1}} 

\def \txvta[#1]{\txv_{{\rm t},#1}} 
\def \rxvta[#1]{\rxv_{{\rm t},#1}} 
\def \txvda[#1]{\txv_{{\rm d},#1}} 
\def \rxvda[#1]{\rxv_{{\rm d},#1}} 
\def \nvba[#1]{\nv_{{\rm d},#1}} 
\def \rxta[#1]{\rx_{{\rm t},#1}} 
\def \txda[#1]{\tx_{{\rm d},#1}} 
\def \rxda[#1]{\rx_{{\rm d},#1}}

\def \ptrve[#1]{\hat{p}_{(\txvte[#1],\rxvte[#1])}}

\def \onev[#1]{\underline{1}_{#1}}
\def \snr {\rho}
\def \snrL {\snr_{\rm L}}
\def \SNR {{\rm SNR}}

\def \MuI {I} 
\def \Ent {H} 
\def \EntR {\cH}
\def \EntInn {{\cH{'}}}

\def \step {\Delta}

\def \E {\mathbb{E}} 

\def \chs {g} 
\def \tn {M} 
\def \rn {N} 
\def \ns  {v}  

\def \nv {\bv} 

\def \Tb {T} 
\def \Tt {{\tau T}} 
\def \Ttone {B} 
\def \Tlim {\Tb} 

\def \sign {\text{sign}}

\DeclarePairedDelimiter{\ceil}{\lceil}{\rceil}

\def \ratio {\alpha}

\def \qh {q_{ \chs}}
\def \qha[#1]{q_{{ \chs},#1}}
\def \Qha[#1]{Q_{{\rm \chs},#1}}

\def \qxa[#1]{q_{{ x},#1}}
\def \Qxa[#1]{Q_{{\rm x},#1}}

\def \bQha[#1]{\bQ_{{\rm h},#1}}
\def \bQxa[#1]{\bQ_{{\rm x},#1}}

\def \snreff {\snr_{\rm {eff}}}

\def \betat {\beta_{\rm{\tau}}}
\def \betad {\beta_{\rm{d}}}

\def \qx {q_{ x}}

\def \qxequ {\bar{q}_{ x}} 
\def \qtilde {\lambda}
\def \qxtilde {\qtilde_x} 
\def \qxtildeequ {\bar{\qtilde}_x} 

\def \qhtilde {\qtilde_{\chs}} 
\def \gainv {\gamma}

\def \Ropt {\cR_{\rm opt}}

\def \dy {\varepsilon}
\def \dx {\delta}

\def \varns {\sigma_{}^2}

\def \snrequ {\bar{\snr}}
\def \varnsequ {\bar{\sigma}_{}^2}

\def \iid {{\it iid}\xspace}
\def \mse {\cE}
\def \msea#1{{\mse}_{#1}}

\def \rxdequa[#1]{\tilde{\rx}_{{\rm d},#1}}

\def \nvequ {\bar{\nv}}
\def \rxvequ {\bar{\rxv}}
\def \rxmequ {\bar{\rxm}}

\def \txvequ {\bar{\txv}}
\def \txequ {\bar{\tx}}

\def \htxta[#1]{\hat{\tx}_{{\rm t},#1}} 
\def \htxda[#1]{\hat{\tx}_{{\rm d},#1}} 
\def \ttxta[#1]{\tilde{\tx}_{{\rm t},#1}} 
\def \ttxda[#1]{\tilde{\tx}_{{\rm d},#1}} 
\def \zta[#1]{z_{{\rm t},#1}}
\def \bzda[#1]{\bz_{{\rm d},#1}}
\def \bzta[#1]{\bz_{{\rm t},#1}}
\def \zda[#1]{z_{{\rm d},#1}}

\def \rxvda[#1]{\rxv_{{\rm d},#1}}

\def \nvq {\nv_{\rm q}}

\def \expeq[#1]{\overset{{#1}}{\equiv}}

\def \Enta[#1]{\Omega_{#1}}
\def \vara[#1]{\sigma^2_{#1}}

\def \IAWGN {I_{\rm AWGN}}

\def \tauopt{\tau_{{\rm opt}}}
\def \upto {{\nearrow}}
\def \downto {{\searrow}}
\def \bit {b}
\def \taup {\tau'}

\def \RL {\cR_{\rm L}}
\def \Rknown {\cR_{\rm known}}

\def \EntInnC#1#2#3#4{\EntInn({#2}_{#4}|{#1}_{#3})}

\def \cIInn#1#2#3{\cI{'}({#1}_{#3};{#2}_{#3})}

\def \hatbar#1{\hat{{\bar{#1}}}}

\def \ptg {p_{\tilde{g}}}
\def \ptx {p_{\tilde{x}}}

\usepackage{amsthm}

\newtheorem{thm}{{Theorem}}

\newtheorem{conject}{Conjecture}

\begin{document}

\title{Training-Based Equivalence Relations in Large-Scale Quantized Communication Systems}
\author{Kang~Gao,~\IEEEmembership{Student~Member,~IEEE,}
Xiangbo~Meng,~\IEEEmembership{Student Member,~IEEE,}
J.~Nicholas~Laneman,~\IEEEmembership{Fellow,~IEEE,} Jonathan~Chisum,~\IEEEmembership{Senior~Member,~IEEE,} Ralf~Bendlin,~\IEEEmembership{Senior~Member,~IEEE,} Aditya~Chopra,~\IEEEmembership{Senior~Member,~IEEE,} and~Bertrand~M.~Hochwald,~\IEEEmembership{Fellow,~IEEE}%
\thanks{Portions of this work were presented in ITA, 2019 (San Diego) and Globecom Workshop, 2019 (Hawaii). This work was generously supported by NSF Grant \#1731056, and AT\&T Labs.}
\thanks{Kang Gao, Xiangbo Meng, J. Nicholas Laneman, Jonathan Chisum, and Bertrand M. Hochwald are with the Department of Electrical Engineering, University of Notre Dame, Notre Dame, IN, 46556 USA (email:kgao@nd.edu; xmeng@nd.edu; jnl@nd.edu; jchisum@nd.edu; bhochwald@nd.edu).}
\thanks{Ralf Bendlin and Aditya Chopra are with AT\&T Labs, Austin, TX, 78712 USA (email:rb691m@att.com; ac116g@att.com).}}
\maketitle
\thispagestyle{plain}
\pagestyle{plain}
\begin{abstract}
We show that a quantized large-scale system with unknown parameters and training signals can be analyzed by examining an equivalent system with known parameters by modifying the signal power and noise variance in a prescribed manner.  Applications to training in wireless communications and signal processing are shown. In wireless communications, we show that the optimal number of training signals can be significantly smaller than the number of transmitting elements.  Similar conclusions can be drawn when considering the symbol error rate in signal processing applications, as long as the number of receiving elements is large enough. We show that a linear analysis of training in a quantized system can be accurate when the thermal noise is high or the system is operating near its saturation rate.
\end{abstract}
\IEEEpeerreviewmaketitle

\section{Introduction}
We consider the system
\begin{align*}
    \rxv_t=\bff\left(\sqrt{\frac{\snr}{\tn}}\chm\txv_t+\nv_t\right), 
    \numberthis
    \label{eq:nonlinear_system_model}
\end{align*}
where $\txv_t$ and $\rxv_t$ are the input and output at time $t$ with dimensions $\tn$ and $\rn$, $\chm$ is an unknown complex $\rn\times \tn$ random matrix whose elements have \iid real and imaginary components with zero-mean half-variance common distribution $\ptg(\cdot)$, and $\txv_1,\txv_2,\ldots$ are independent of $\chm$, and their elements have \iid real and imaginary components with zero-mean half-variance common distribution $\ptx(\cdot)$.  The elements of $\nv_t$ are \iid circular-symmetric complex Gaussian $\cC\cN(0,\varns)$, independent of the input and $\chm$, $\bff(\cdot)$ is an element-wise function that applies $\bit$-bit uniform quantization $f(\cdot)$ to each element, and the real and imaginary parts are quantized independently.  When $\varns=1$, the quantity $\snr$ is nominally called the signal-to-noise ratio (SNR) because it represents the ratio of the average signal energy $(\snr/\tn)\E\|G\txv_t\|^2=\snr\rn$ to noise variance $\rn\varns=\rn$, before quantization.

The quantizer $f(\cdot)$ with $\bit$ bits has $2^\bit-1$ real quantization thresholds defined as
\begin{align}
    r_k=(-2^{\bit-1}+k)\step,\; {\rm{for }}\; k=1,2,\cdots,2^\bit-1,
    \label{eq:threshold_levels}
\end{align}
where $\step$ is the quantization step size. We define $r_0=-\infty$, and $r_{2^\bit}=+\infty$
for convenience. The output of the quantizer indicates the quantization level: $f(w)=k$ for $w\in(r_{k-1},r_k]$ and $k=1,\ldots,2^\bit$. We use $\bit=\infty$ to denote the case when $f(w)=w$ for $w\in\cR$ (quantizer is removed). When the input to the quantizer is a complex number, its real and imaginary parts are quantized independently.  Often, $\step$ is designed as a function of $\bit$ to make full use of each quantization level. It is assumed throughout our numerical results that $\step$ is chosen such that $f(w)=1$ or $f(w)=2^b$ with probability $1/2^b$ when the input distribution on $w$ is real Gaussian with mean zero and variance $(1+\snr)/2$.
For example, when $\bit=2$, we have $\step=0.47\sqrt{\snr+1}$ and when $\bit=3$, we have $\step=0.27\sqrt{\snr+1}$.  However, the main trends and conclusions contained herein are not sensitive to these choices.

The system has a ``blocklength", denoted as $\Tb$, during which the unknown matrix $\chm$ is constant, and after which it changes independently to a new value. It is desired to send known training signals from which information about $\chm$ can be learned from the $(\txv_t,\rxv_t)$ input-output pairs.  Note that the nonlinearity $f(\cdot)$ may make it difficult to obtain accurate information about $\chm$ during training, but it is also conceivable that only limited information about $\chm$ is needed, depending on the desired application of the model \eqref{eq:nonlinear_system_model}.

The model \eqref{eq:nonlinear_system_model} is widely used in wireless communications and signal processing models \cite{hassibi2003much,li2016much,li2017channel,takeuchi2010achievable,takeuchi2012large,takeuchi2013achievable,wen2015performance,wen2015joint,wen2016bayes,estes2020efficient}, where $\txv_t$ and $\rxv_t$ are the transmitted and received signals at time $t$ in a multiple-input-multiple-output (MIMO) system with $\tn$ transmitters and $\rn$ receivers, $\chm$ models the channel coefficients between the transmitters and receivers, $\Tb$ is the (integer) coherence time of the channel in symbols, $\nv_t$ is the additive noise at time $t$, $f(\cdot)$ models receiver effects such as quantization in analog-to-digital converters (ADC's) and nonlinearities in amplifiers.  
For example, single-bit ADC's with $f(x)=\sign(x)$ are considered in \cite{li2016much,li2017channel}, and low-resolution ADC's with uniform quantizers are considered in \cite{wen2015joint,wen2016bayes}. Part of the coherence time is typically used for training to learn $\chm$, while the remainder is used for data transmission or symbol detection. So-called ``one-shot'' learning is considered in \cite{hassibi2003much,li2016much,li2017channel} where $\chm$ is learned only from training, while \cite{takeuchi2010achievable,takeuchi2012large,takeuchi2013achievable,wen2015performance,wen2015joint,wen2016bayes} allows $\chm$ to be refined after training. 
Often, linearization of $f(\cdot)$ is used to aid the analysis \cite{li2016much,li2017channel}.  We are primarily interested in one-shot learning in the limit $\Tb\to\infty$, when closed-form analysis is possible.

Using a training-based lower bound on mutual information for large-scale systems \cite{gaopart1}, we show that \eqref{eq:nonlinear_system_model} can be analyzed by examining an equivalent system with known parameters by modifying the signal power and noise variance in a prescribed manner.  We show that the number of training signals can be significantly smaller than the number of transmitting elements in both wireless communication and signal processing applications. We show that a linear analysis of \eqref{eq:nonlinear_system_model} can be accurate when the thermal noise is high or the system is operating near its saturation rate.

\section{Equivalence in entropy and mutual information}

Let $\Ttone = \ceil{\Tt}$ be the training time where $\tau\in(0,1]$ is the fraction of the blocklength used to learn $\chm$. 
We allow $\rn$ and $\tn$ to increase proportionally to the blocklength $\Tb$ as $\Tb\to\infty$. 
The ratios are
\begin{align}
    \ratio=\frac{\rn}{\tn},\quad \beta=\frac{\Tb}{\tn}.
    \label{eq:ratios}
\end{align}

Define $\txm_t=[\txv_1,\txv_2,\cdots,\txv_t]$, $\rxm_t=[\rxv_1,\rxv_2,\cdots,\rxv_t]$, and
\begingroup
\begin{align*}
    \cIInn{\txA}{\rxA}{}&=\lim_{\Tlim\to\infty}\frac{1}{\rn}\MuI(\txv_{\Ttone+1};\rxv_{\Ttone+1}|\txm_{\Ttone},\rxm_{\Ttone}),
    \numberthis
    \label{eq:MuI_def_high}
\\
\EntInnC{\txA}{\rxA}{^+}{}&=\lim_{\Tlim\to\infty}\frac{1}{\rn}\Ent(\rxv_{\Ttone+1}|\txm_{\Ttone+1},\rxm_{\Ttone}),
    \numberthis
    \label{eq:EntInn_cond_def_vec}
\\
\EntInnC{\txA}{\rxA}{}{}&=\lim_{\Tlim\to\infty}\frac{1}{\rn}\Ent(\rxv_{\Ttone+1}|\txm_{\Ttone},\rxm_{\Ttone}).
    \numberthis
    \label{eq:EntInn_def_vec}
\end{align*}
\endgroup
The quantities \eqref{eq:MuI_def_high}--\eqref{eq:EntInn_def_vec} are shown in \cite{gaopart1} to play an important role in determining the optimum amount of training in a system with unknown parameters.  We show that these quantities equal those of another ``equivalent system'' where $\chm$ is known
, which is stated in the following theorem.
\begin{thm}
\label{thm:equivalence_in_Ent}
For the system \eqref{eq:nonlinear_system_model} with $\Ttone$ input-output training pairs $(\txm_{\Ttone},\rxm_{\Ttone})$,
\begingroup
\begin{align}
    \cIInn{\txA}{\rxA}{} = \lim_{\Tlim\to\infty}\frac{1}{\rn}\MuI(\txvequ;\rxvequ|\chm),
    \label{eq:equi_in_MuI}
\\
    \EntInnC{\txA}{\rxA}{ ^+}{} = \lim_{\Tlim\to\infty}\frac{1}{\rn}\Ent(\rxvequ|\chm,\txvequ),
    \label{eq:equi_in_Ent_Y_cond_X}
\\
    \EntInnC{\txA}{\rxA}{}{} = \lim_{\Tlim\to\infty}\frac{1}{\rn}\Ent(\rxvequ|\chm).
    \label{eq:equi_in_Ent_Y}
\end{align}
\endgroup
where the mutual information and entropies of the right-hand sides of the above equations are derived from the system
\begin{align}
    \rxvequ=\bff\left(\sqrt{{\snrequ}/{\tn}}\chm\txvequ+\nvequ\right),
    \label{eq:equivalent_system_thm}
\end{align}
where $\chm$ is known at the receiver, $\txvequ$ is distributed identically to $\txv_t$, the entries of $\nvequ$ are \iid $\cC\cN(0,\varnsequ)$,
and $\snrequ,\varnsequ$ are defined as
\begin{align}
      \snrequ = \snr(1-\msea{\chm}),\qquad\varnsequ = \varns + \snr\cdot\msea{\chm},
      \label{eq:SNR_var_equivalent}
\end{align}
where $\msea{\chm}$ is
\begin{align*}
    \msea{\chm}=\lim_{T\to\infty}\frac{1}{\tn\rn}\E\norm{\chm-\hat{\chm}}^2_{\rm F},
    \numberthis
    \label{eq:MSE_MMSE_G}
\end{align*}
with $\hat{\chm} = \E(\chm|\txm_{\Ttone},\rxm_{\Ttone}).$
\end{thm}
\noindent
{\bf Proof:} Please see Appendix \ref{app:proof_equivalence}. 

The theorem is similar to some well-known results for the linear-system model $\rxv_t=\sqrt{\frac{\snr}{\tn}}\chm\txv_t+\nv_t$ (which omits the $\bff(\cdot)$ function in \eqref{eq:nonlinear_system_model}), where the unknown $\chm$ is replaced by its minimum mean-square error (MMSE) estimate obtained from the training signals, and the system is converted to one where $\chm$ is known, and the estimation error is converted to Gaussian noise that is added to $\nv_t$.  Generally, these existing results are in the form of lower bounds on mutual information that come from approximating the estimation error as (worst-case) Gaussian noise that is independent of $\nv_t$  \cite{hassibi2003much,medard2000effect}.  However, there are some key differences in Theorem \ref{thm:equivalence_in_Ent}: (i) The theorem applies in the large-scale system limit, and provides exact equalities, not just lower bounds; (ii) As a result of the large-scale limit, the model \eqref{eq:nonlinear_system_model} does not require Gaussian assumptions on $\chm$ or $\txv_t$, worst-case noise analysis, or any linearization of the quantizer $f(\cdot)$.

This theorem is useful because quantities such as the entropies and mutual informations for the model \eqref{eq:nonlinear_system_model} with known $\chm$ are generally easier to compute than those for unknown $\chm$, and the effect of the unknown $\chm$ is converted to the parameter $\msea{\chm}$, which is the large-scale limit of MSE in the MMSE estimate of $\chm$.  This is of value in Sections \ref{sec:wireless_communication}--\ref{sec:signal_processing}, where known-$\chm$ results are leveraged to obtain results in communication and signal processing problems where $\chm$ is estimated through training.  
Although we compute large-scale limits, it is anticipated that the results herein provide good approximations for systems with finite $\tn$, $\rn$, and $\Tb$ simply by substituting the $\ratio$ and $\beta$ computed for the finite-dimensional system into the limiting formulas.  Evidence that this approximation is reasonably accurate for systems of moderate dimensions is given in Section \ref{sec:finite_OK}. 

The following steps summarize the computations needed in the theorem.

\subsection{Computing $\cIInn{\txA}{\rxA}{}$}
\label{sec:step_by_step_computation}
\begin{Process}[H]
\SetAlgoLined
\nonl\algorithmicrequire{  $b$,
$r_k$ \eqref{eq:threshold_levels}, 
$\ptx(\cdot)$, 
$\ptg(\cdot)$, 
$\tau, \alpha, \beta, \varns, \snr$, 
$\IAWGN(\cdot,\cdot)$ and $\mse(\cdot,\cdot)$ \eqref{eq:MuI_AWGN_def}--\eqref{eq:p_wy_joint},
$\Enta[](\gainv,s)$ and $\chi(\gainv,s)$ \eqref{eq:Ent_V_o_def}--\eqref{eq:psi_prime_b} or \eqref{eq:linear_Ent_chi}}
 (see Appendix \ref{app:proof_equivalence})\;
\nonl\algorithmicensure{ $\cIInn{\txA}{\rxA}{}$ \eqref{eq:MuI_def_high}
}\;
Solve $\msea{\chm}$ defined in \eqref{eq:MSE_MMSE_G} from \eqref{eq:qh_qhtilde_solu_A&mse_G_vs_qh}\;
Compute $\snrequ$ and $\varnsequ$ from \eqref{eq:SNR_var_equivalent}\;
Compute \eqref{eq:MuI_know_G}, and use \eqref{eq:equi_in_MuI}.
 \caption{Compute $\cIInn{\txA}{\rxA}{}$}
\end{Process}

Further details of the computations appear in Appendix \ref{app:proof_equivalence} as part of the proof of Theorem \ref{thm:equivalence_in_Ent}.
The theorem and computations also hold for systems with real $\chm,\txv_t,$ and $\nv_t$, when both $\chm$ and $\txv_t$ consist of \iid elements with zero mean and unit variance, and the elements of $\nv_t$ are \iid $\cN(0,1)$.   However, the steps above require minor modifications. First, the $\ptg(\cdot)$ and $\ptx(\cdot)$ as used in  \eqref{eq:qh_qhtilde_solu_A&mse_G_vs_qh}, \eqref{eq:qx_solu_known_G}, \eqref{eq:entropy_yt_known_G} and \eqref{eq:MuI_know_G} should be the distributions of the elements of $\chm$ and $\txv_t$ normalized by $\frac{1}{\sqrt{2}}$ to obtain variance equal to $1/2$. Second, the $r_k$ as used in \eqref{eq:psi_func} and \eqref{eq:psi_prime_b} should be the actual $r_k$ divided by $\sqrt{2}$. 
Finally, the actual values of $\cIInn{\txA}{\rxA}{}, \EntInnC{\txA}{\rxA}{ ^+}{},$ and $\EntInnC{\txA}{\rxA}{}{}$ are computed by \eqref{eq:entropy_yt_known_G}--\eqref{eq:MuI_know_G} and then dividing by two.

\section{Application to Wireless Communication}
\label{sec:wireless_communication}
In communication systems we are often interested in maximizing the achievable rate. We consider a MIMO system \cite{wen2016bayes,larsson2014massive,lu2014overview,marzetta2016fundamentals,bjornson2019massive} modeled by
 \eqref{eq:nonlinear_system_model}, where $\txv_t$ models the transmitted signals from $\tn$ elements (transmitter antennas) at time $t$, $\rxv_t$ models the received signals with $\rn$ elements (receiver antennas), $\chm$ models the unknown baseband-equivalent wireless channel, $\nv_t$ models the additive white Gaussian noise at the receiver, $f(\cdot)$ models the uniform $\bit$-bit quantization at the analog-to-digital converters, $\Tb$ models the coherent blocklength during which the channel is constant, and a fraction $\tau$ of the total blocklength is used for training to learn the channel. In a Rayleigh environment, $\chm$ has \iid $\cC\cN(0,1)$ elements, and the corresponding $\ptg(\cdot)$ is $\cN(0,\frac{1}{2})$.  We assume that the real and imaginary elements of the transmitted vector $\txv_t$ are \iid with zero mean and half variance and are generated using $\txbit$-bit uniform digital-to-analog converters (DAC's) in both the in-phase and the quadrature branches.  This creates a $2^{2\txbit}$-QAM constellation, with all possible symbols generated with equal probability; the corresponding $\ptx(\cdot)$ is uniform among the $2^\txbit$ real and imaginary components.  We use $\txbit=\infty$ to denote an unquantized transmitter where the elements of $\txv_t$ are \iid $\cC\cN(0,1)$ and the corresponding $\ptx(\cdot)$ is $\cN(0,\frac{1}{2})$.  Throughout this section we assume $\varns=1$.

Using the results in \cite{gaopart1}, we conclude that the optimal achievable rate (in ``bits/channel-use/transmitter") for a trained system is
\begin{align}
    \Ropt=\max_{\tau} (1-\tau)\ratio\cIInn{\txA}{\rxA}{},
    \label{eq:opt_rate_per_tx}
\end{align}
where $\cIInn{\txA}{\rxA}{}$ is defined in \eqref{eq:MuI_def_high}. 
The optimal training fraction is
\begin{align}
    \tauopt=\argmax_{\tau}(1-\tau)\ratio\cIInn{\txA}{\rxA}{}.
    \label{eq:tau_opt_value}
\end{align}
For comparison, we sometimes compute the rate per transmitter for systems with known $\chm$, 
\begin{align}
    \Rknown=\lim_{\tn\to\infty}\MuI(\txv_t;\rxv_t|\chm)/{\tn},
    \label{eq:Rknown}
\end{align}
which can be computed from \eqref{eq:MuI_know_G} with $(\snrequ,\varnsequ)$ replaced by $(\snr,1)$. Since $\txv_t$ are \iid, \eqref{eq:Rknown} is not a function of $t$.  

The parameter $\beta=\Tb/\tn$ is the ratio of the coherence time of the channel (in symbols) to the number of transmitters and is therefore strongly dependent on the physical environment.  We may choose a typical value as follows: Suppose we choose a 3.5 GHz carrier frequency with maximum mobility of 80 miles/hour; the maximum Doppler shift becomes $f_{\rm d}=\frac{80\; \text{miles/h}\times 3.5\; \text{GHz}}{3\times 10^8\; \text{m/s}}=417$ Hz, and the corresponding coherence time is  $\frac{9}{16\pi f_{\rm d}}=0.4$ ms \cite{rappaport1996wireless}. We consider 10 MHz bandwidth and assume that the system is operated at Nyquist sampling rate (10 complex Msamples/second), which produces $\Tb=4000$ discrete samples during each 0.4 ms coherent block. In a system with $\tn=100$ elements at the transmitter, we obtain $\beta=40$.  The remainder of this section considers the results of \eqref{eq:opt_rate_per_tx}--\eqref{eq:Rknown} for various scenarios.  Details can be found in the figure captions.

\subsubsection{More receivers can compensate for lack of channel information}
\begin{figure}
\includegraphics[width=3.5in]{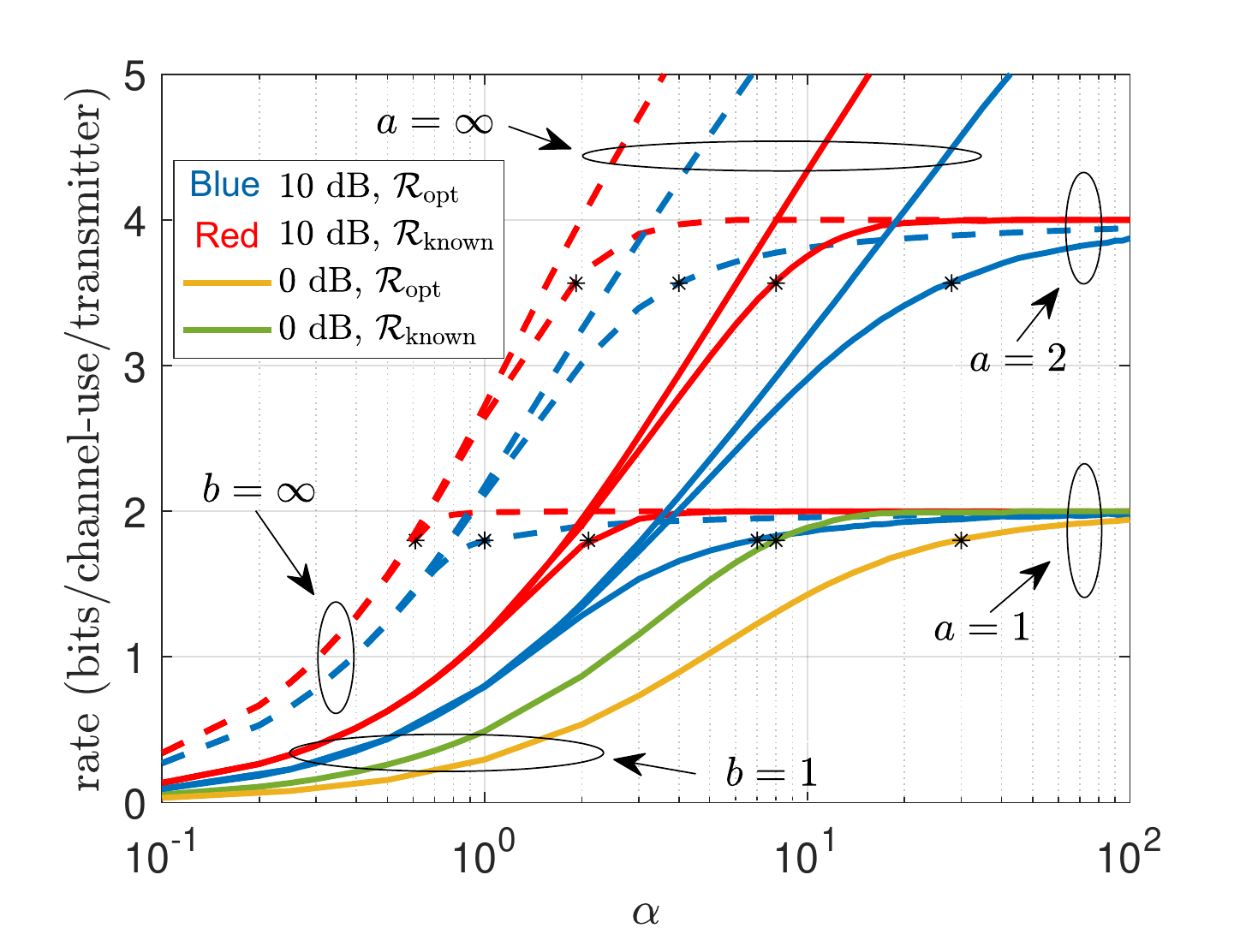}
\centering
    \caption{Solutions of \eqref{eq:opt_rate_per_tx} showing $\Ropt$ vs $\ratio$ for $\txbit=1,2,\infty$, and $\bit=1$ (solid curves) and $\bit=\infty$ (dashed curves), and SNR $=0,\,10$ dB, for $\beta=40$ (see text for explanation of this choice).  Also shown is $\Rknown$ \eqref{eq:Rknown}.  The curves saturate at $2$ for $\txbit=1$, and at $4$ for $\txbit=2$, while there is no saturation for $\txbit=\infty$.    Note by comparing $\Ropt$ (unknown channel) versus $\Rknown$ (known channel) at the 90\% level (indicated by ``*") that the price in $\alpha$ for not knowing the channel is higher when comparing linear ($b=\infty$) versus one-bit ($b=1$) receivers.  For example, with $\txbit=2$, observe that the ``*" on the dashed-red curve and dashed-blue curve are at $\ratio=2$ and $\ratio=4$, respectively; on the other hand, the ``*" on the solid-red curve and solid-blue curve are at $\ratio=8$ and $\ratio=28$, indicating that $\alpha$ has to increase more when the receiver resolution is lower to compensate for lack of channel information.
    Observe also that, generally, increasing $\alpha$ compensates for lack of resolution at the receiver.  
    This effect is independent of whether the channel is known or trained at the receiver.  The asymptotes $\Ropt=2$ and $\Ropt=4$ are reached quite slowly when the channel is unknown.}
    \label{fig:alpha_saturation_one_bit}
\end{figure}

In Fig. \ref{fig:alpha_saturation_one_bit} we consider rate versus $\ratio$ for $\txbit=1,2$; the maximum rates per transmitter are then 2 bits and 4 bits respectively. These asymptotes are approached as $\ratio$ is increased, but are reached much more slowly when the channel is unknown than when it is known, as seen by comparing blue curves versus the corresponding red curves, or the yellow curve versus the green curve.  Larger $\ratio=\rn/\tn$ represents larger number of receivers per transmitter.  The linear receiver ($\bit=\infty$, dashed curves) and the one-bit quantized receiver ($\bit=1$, solid curves), and the linear transmitter ($\txbit=\infty$) are shown for comparison.

\subsubsection{Very limited channel information is sometimes sufficient}
\begin{figure}
\includegraphics[width=3.5in]{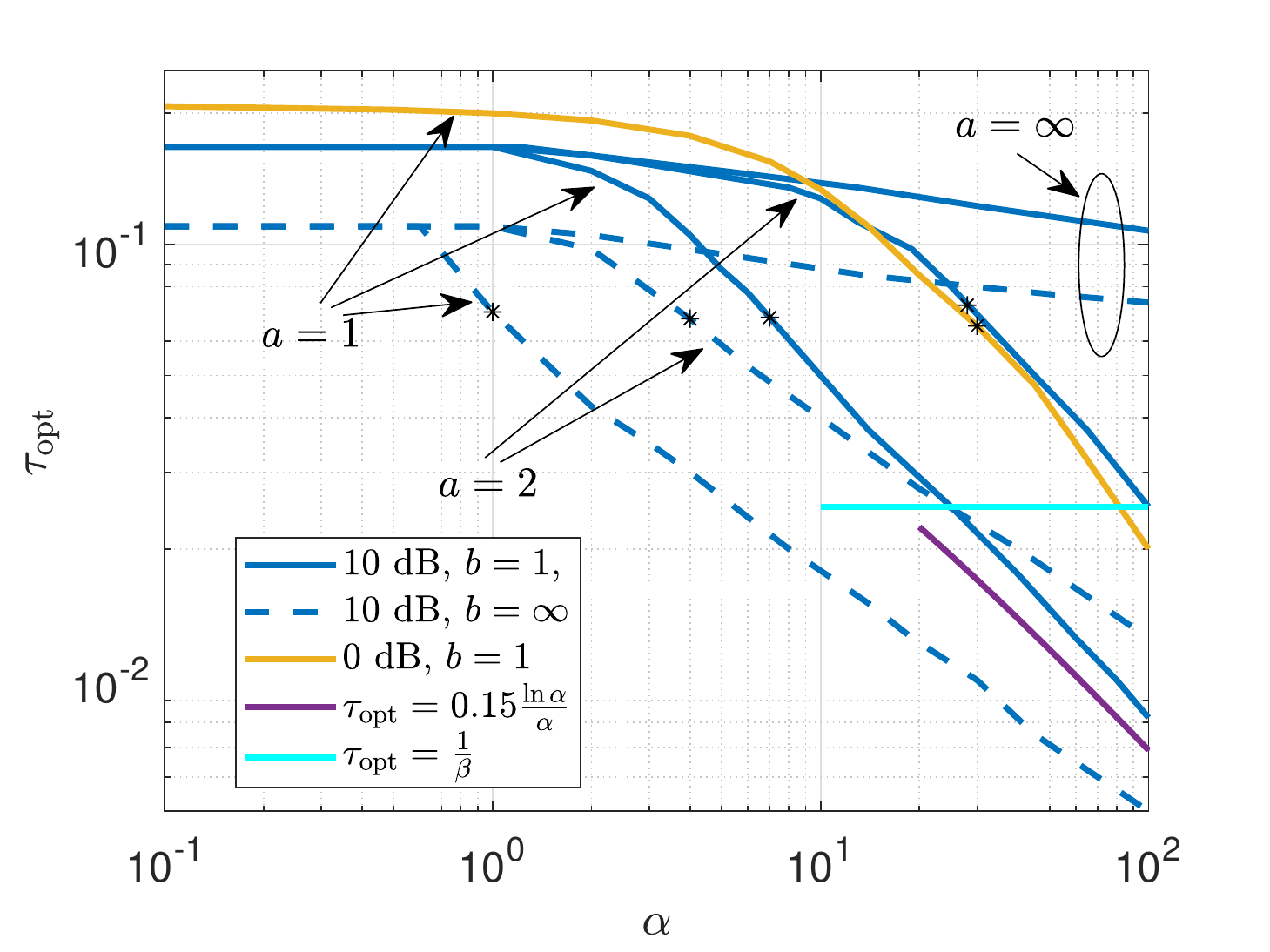}
\centering
    \caption{Solutions of \eqref{eq:tau_opt_value} showing $\tauopt$ vs $\ratio$, where the 90\% levels are marked as in Fig.\ \ref{fig:alpha_saturation_one_bit} (omitting known-channel results). Note that $\tauopt$ is generally insensitive to $\ratio$ when $\ratio$ is small, and decreases rapidly with $\ratio$ as $\Ropt$ approaches the saturation rate $2\txbit$.  The markers suggest that $\tauopt\approx 0.07$ independently of the $\txbit$ (transmitter resolution), SNR, or $\bit$ (receiver resolution).  As $\ratio$ grows, eventually $\tauopt\cdot\beta<1$ (indicated by the solid cyan line), at which point the number of training symbols is smaller than the number of transmitter elements.  Also shown in purple is the large-$\alpha$ result (\ref{eq:tauopt_one_bit_output}) for 10 dB SNR, $\txbit=1$ and $\bit=1$.}
    \label{fig:beta_40_tau_opt_smooth}
\end{figure}

In Fig. \ref{fig:beta_40_tau_opt_smooth}, we show that for $\txbit=1,2$, very limited channel information is needed when $\ratio$ is large because the corresponding $\tauopt$ is small. The optimum number of training signals may be smaller than the number of transmitters ($\tauopt\beta<1$).  This is also shown to a limited extent for $\txbit=1$ and $\bit=1$ in \cite{kang2019training,gao2019channel}.  We show in Section \ref{subsub:tau_vs_alpha} that $\tauopt$ decays as $(\ln\alpha)/\alpha$ for large $\ratio$ when $\txbit=1$.

\subsubsection{Quantization effects limit how small $\alpha$ can be}
\begin{figure}
\includegraphics[width=3.5in]{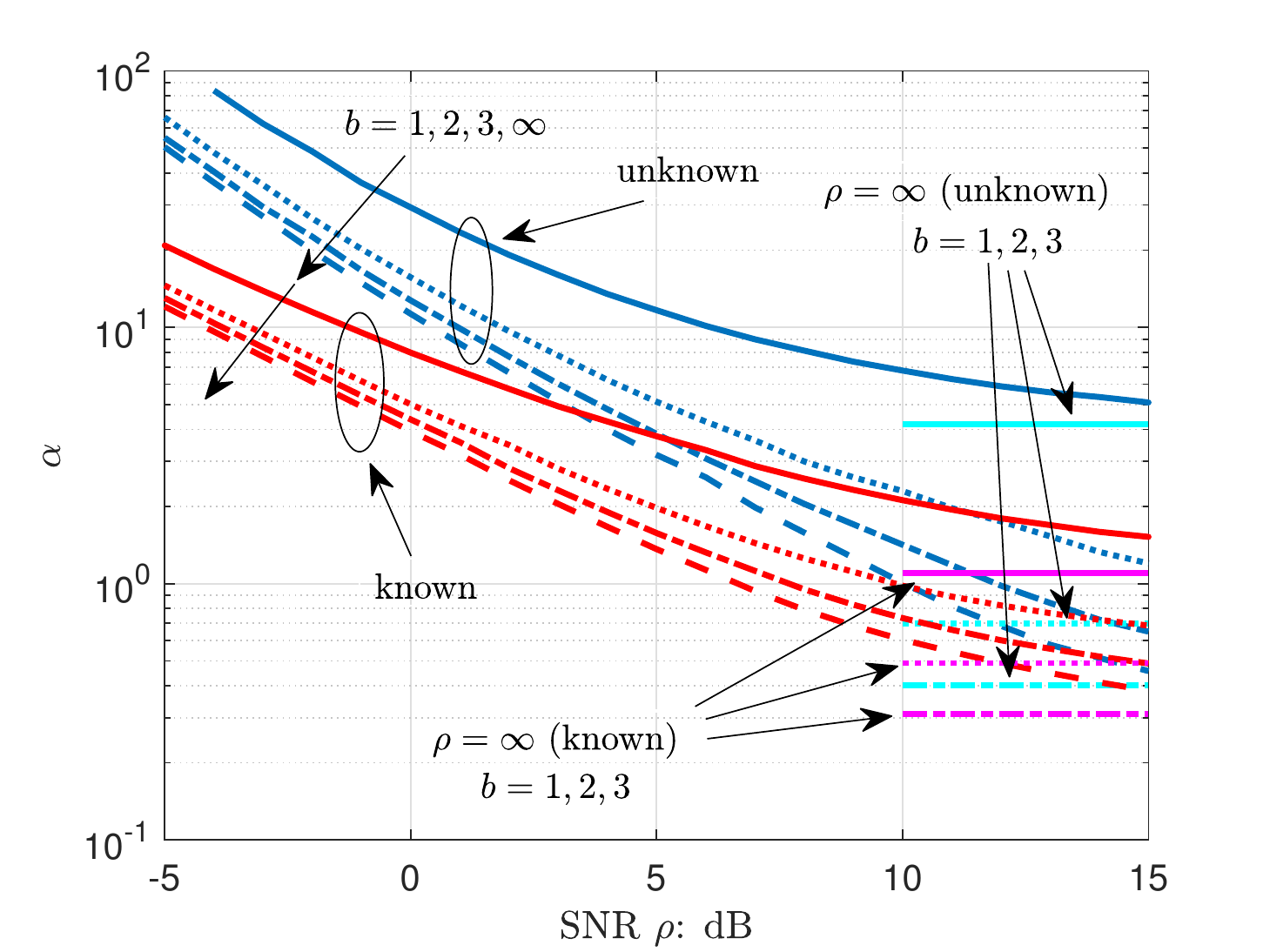}
\centering
    \caption{Solutions of \eqref{eq:opt_rate_per_tx} showing $\ratio$ versus SNR when $\Ropt=1.8$ (90\% level), $\beta=40$, $\txbit=1$; also shown are known-channel solutions of \eqref{eq:Rknown}.  The curves decrease as SNR increases and reach asymptotes as $\snr=\infty$ that are shown in cyan (unknown channel) and magenta (known channel) with $\bit=1,2,$ and 3.  These asymptotes are the result of the quantization noise at the receiver that restrict the $\alpha$ from going to zero as $\snr=\infty$.  The unknown-channel asymptotes are above the known-channel asymptotes because the channel is estimated through the quantized receiver, creating extra effective noise that needs larger $\alpha$ to compensate.  For linear receivers ($\bit=\infty$), perfect channel estimation can be obtained as $\snr=\infty$, and therefore there are no asymptotes.  At low SNR, the slopes of the red curves are similar to each other (as are the blue curves) because the additive (thermal) noise dominates the quantization noise, and the effect of quantization can be treated as degradation in SNR that depends on $\bit$.
    }
    \label{fig:fixed_tau_opt_SNR_vs_alpha_2}
\end{figure}

The values of $\ratio$ required to achieve $\Ropt=1.8$ ($90\%$ level) for various SNR $\snr$ and $\bit$ with $\txbit=1$ are shown in Fig. \ref{fig:fixed_tau_opt_SNR_vs_alpha_2}. It is clear that $\ratio$ decreases as $\snr$ increases, and there are asymptotes when $\snr=\infty$ for $\bit=1,2,3$ whether the channel is known or not, because of the quantization noise.  When $\bit=\infty$, there are no asymptotes since the channel can be estimated perfectly, and the discrete transmitted signal can be detected perfectly as $\snr\rightarrow\infty$.

\subsubsection{Linearization works well near the saturation rate or with high thermal noise}

Linearizing the system model \eqref{eq:nonlinear_system_model} at the receiver allows us to model the quantization noise in a variety of ways. 
For example, when $\bit=1$, we assume that the real and imaginary parts of the received signal are taken from $\{\frac{\pm 1}{\sqrt{2}}\}$.  For $\bit=2$, we assume that they are taken from $\{(\frac{\pm 1}{\sqrt{10}},\frac{\pm 3}{\sqrt{10}})\}$. These values are selected so that the output of $f(\cdot)$ has zero mean and unit variance.

By using the Bussgang decomposition \cite{bussgang1952crosscorrelation,jacobsson2017throughput}, we can reformulate the system in \eqref{eq:nonlinear_system_model} as 
\begin{align*}
    \rxv_t=\sqrt{\frac{\eta}{(\snr+1)}}\left(\sqrt{\frac{\snr}{\tn}}\chm\txv_t+\nv_t\right)+\nvq,
    \numberthis
    \label{eq:AQN_model_buss}
\end{align*}
where $\nvq$ is uncorrelated with $\sqrt{\frac{\snr}{\tn}}\chm\txv_t+\nv_t$, $\nvq$ has zero mean with covariance matrix $(1-\eta)I$, and where $\eta=2/\pi$
for $\bit=1$, and $\eta = \frac{2}{5\pi}(1+2e^{-\frac{\step^2}{\snr+1}})^2$ for $\bit=2$.

For tractability, we assume that $\nvq\sim\cC\cN(0,(1-\eta)I)$ and is independent of $\chm, \txv_t,$ and $\nv_t$. Then,  \eqref{eq:AQN_model_buss} can be considered as a system with SNR $\snrL$:
\begin{align}
    \rxv_t=\sqrt{{\snrL}/{\tn}}\chm\txv_t+\nv_t,
    \numberthis
    \label{eq:equivalent_linear_model}
\end{align}
where $\snrL = \frac{\eta\frac{\snr}{\snr+1}}{\eta\frac{1}{\snr+1}+(1-\eta)}=\frac{\eta\snr}{(1-\eta)\snr+1}.$
It is shown in \cite{hassibi2003much} that orthogonal training minimizes the mean-square error (MSE) for estimating the channel in \eqref{eq:equivalent_linear_model}. The classical treatment of this model treats the estimated channel as the ``true" channel, while the estimation error is treated as additive Gaussian noise, thereby obtaining a capacity lower bound for any $\txbit$.  We thereby obtain
\begin{align}
    \bar{\rxv}_t = \sqrt{{\snreff}/{\tn}}{\bar{\chm}}\txv_t+\bar{\nv}_t,
    \label{eq:effective_linear_SNR}
\end{align}
where $\snreff$ is the effective SNR
\begin{align}
    \snreff=\frac{\tau\beta\snrL^2}{1+(1+\tau\beta)\snrL},
    \label{eq:effective_SNR}
\end{align}
$\bar{\chm}$ is the estimated channel whose elements are \iid $\cC\cN(0,1)$, and $\bar{\nv}_t$ has \iid $\cC\cN(0,1)$ elements. This known-$\chm$ model has achievable rate 
\begin{align*}
    \cR_{\rm eff} = \lim_{\tn\to\infty}\frac{1}{\tn}\MuI(\txv_t;\bar{\rxv}_t|\bar{\chm}),
\end{align*}
which can be computed using \eqref{eq:MuI_know_G}. Note that $\cR_{\rm eff}$ is a function of $\tau$, since $\snreff$ in \eqref{eq:effective_SNR} is a function of $\tau$. We then define
\begin{align}
    \RL=\max_{\tau}(1-\tau)\cR_{\rm eff}.
    \label{eq:linearization_lower_bound}
\end{align}

\begin{figure}
\includegraphics[width=3.5in]{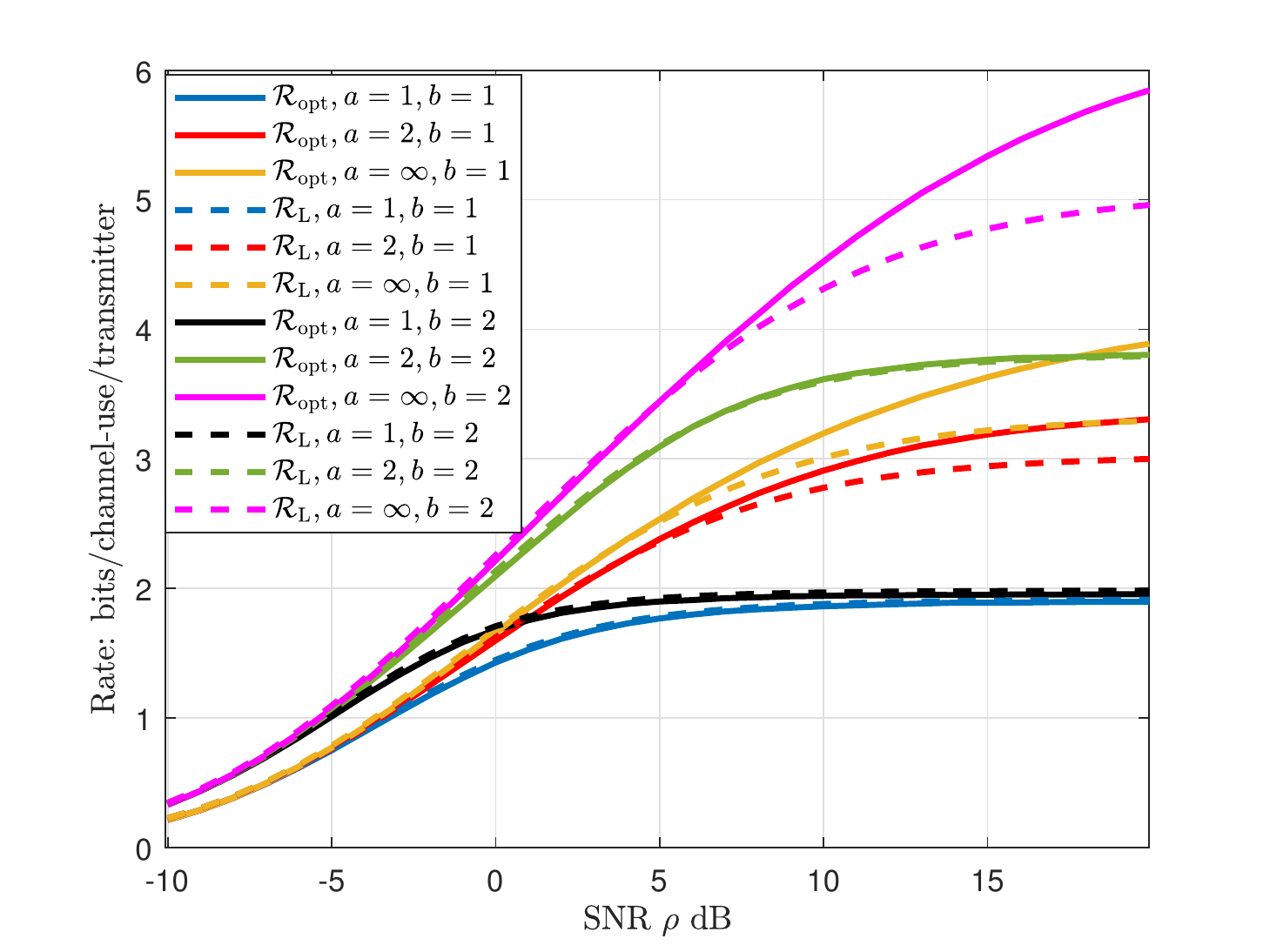}
\centering
    \caption{Solutions of \eqref{eq:opt_rate_per_tx} and \eqref{eq:linearization_lower_bound} showing $\Ropt$ and $\RL$ vs SNR with $\beta=40$ and $\ratio=10$ for $\txbit=1,2,\infty$ and $\bit=1,2$. Observe that $\RL$ is a good approximation of $\Ropt$ below 6 dB SNR, and is sometimes also a good approximation for all $\SNR$, depending on where saturation (rate $2\txbit)$ is reached.}
    \label{fig:compare_with_Bussgang}
\end{figure}

\begin{figure}
\includegraphics[width=3.5in]{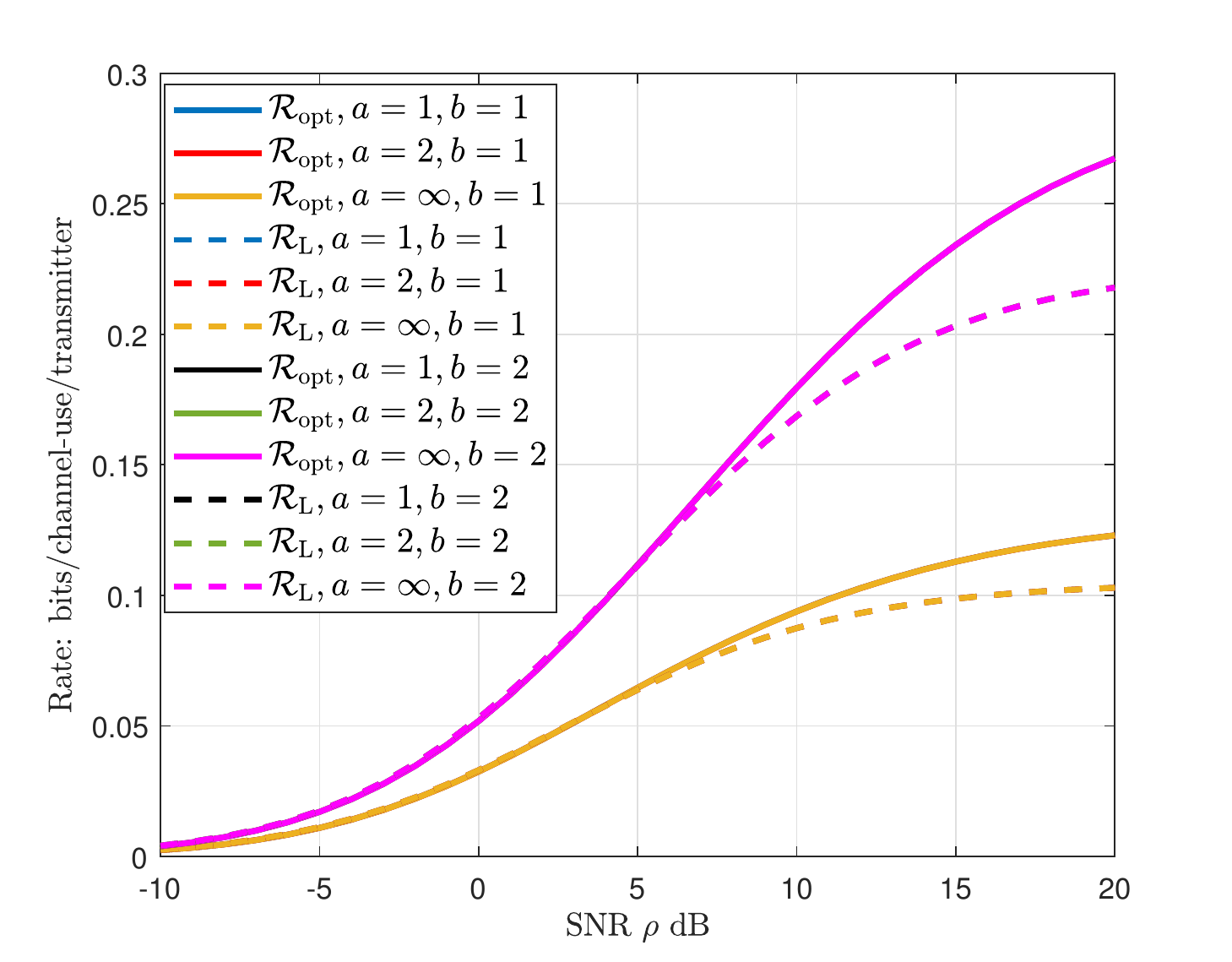}
\centering
    \caption{Solutions of \eqref{eq:opt_rate_per_tx} and \eqref{eq:linearization_lower_bound} for $\Ropt$ and $\RL$ vs SNR with $\beta=40$ and $\ratio=0.1$ for $\txbit=1,2,\infty$ and $\bit=1,2$. Note that the blue, red, and yellow curves that are solid or dashed are on top of each other, while the black, green, and magenta curves that are solid or dashed are on top of each other. This indicates that
    for small $\ratio$ ($\ratio=0.1$), $\Ropt$ and $\RL$ are not sensitive to $\txbit$. Also, $\RL$ is a good approximation of $\Ropt$ when the SNR is below 6 dB.}
    \label{fig:RL_vs_Ropt_alpha_0.1}
\end{figure}

The path just described to obtain $\RL$ involves several approximations, and hence it is unclear how closely $\RL$ should follow $\Ropt$.
However, a comparison between $\Ropt$ \eqref{eq:opt_rate_per_tx} and $\RL$ \eqref{eq:linearization_lower_bound} with $\txbit=1, 2, \infty$ and $\bit=1, 2$ for $\beta= 40$ is shown in Fig. \ref{fig:compare_with_Bussgang} with $\ratio=10$ and in Fig. \ref{fig:RL_vs_Ropt_alpha_0.1} with $\ratio=0.1$. In both Fig. \ref{fig:compare_with_Bussgang} and Fig. \ref{fig:RL_vs_Ropt_alpha_0.1}, we see that $\RL$ is generally a good approximation of $\Ropt$ when the SNR is below 6 dB, but is also accurate above 6 dB in cases where $\Ropt\approx 2\txbit$ (saturation rate) when $\SNR\approx 6$ dB; see especially the blue, black, and green curves in Fig. \ref{fig:compare_with_Bussgang}. Thus at low SNR (high thermal noise) or when the rates are near saturation at low SNR, we can use 
the linear analysis to approximate $\Ropt$.  We also observe that both $\Ropt$ and $\RL$ are not sensitive to $\txbit$ when $\ratio=0.1$.  More discussion of $\Ropt$ with small $\ratio$ is shown in Section \ref{subsec:small_ratio_Ropt}).
\subsubsection{The ratio $\ratio$ is sensitive to $\beta$ when $\beta$ is small}
\begin{figure}
\includegraphics[width=3.5in]{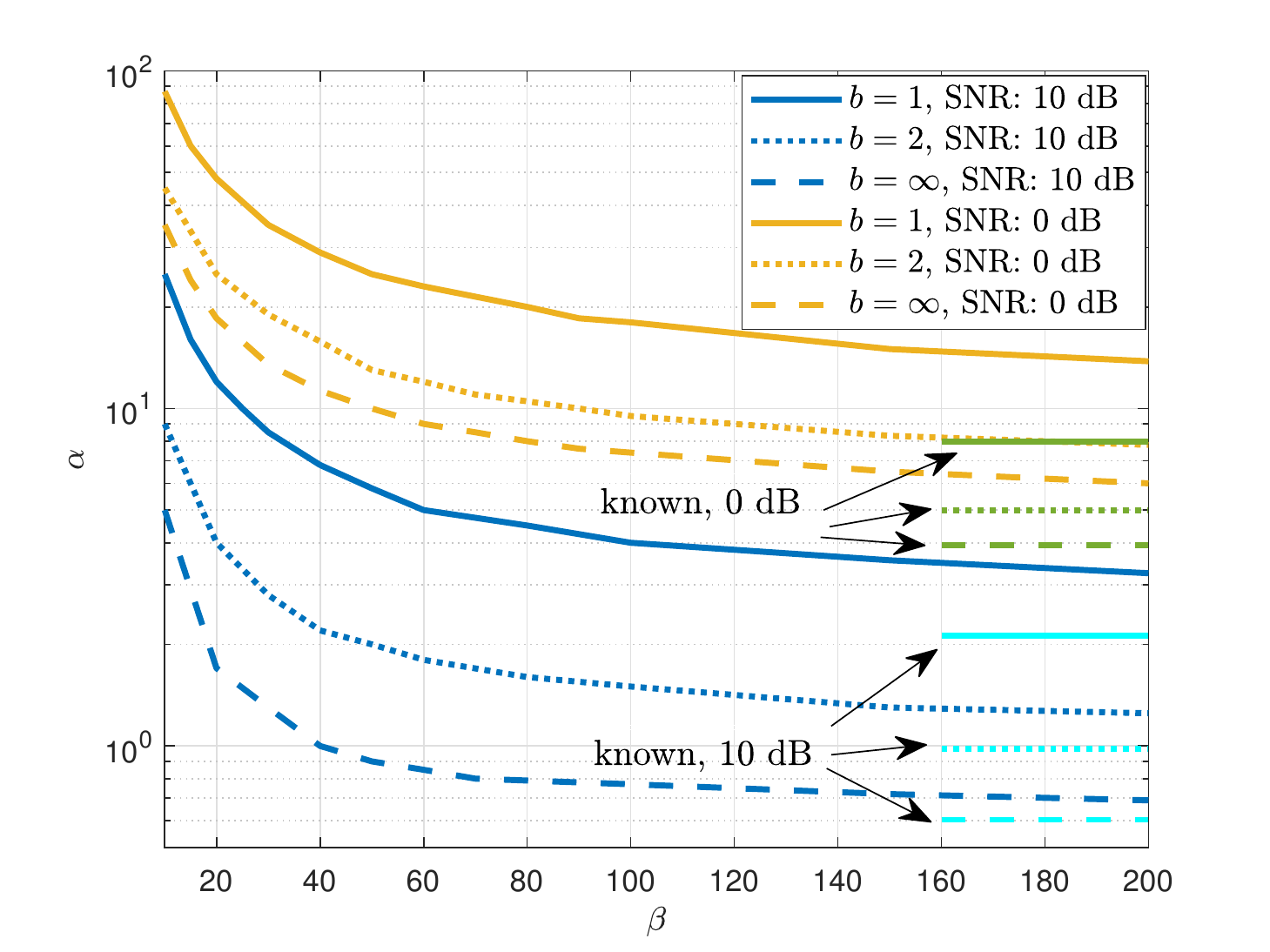}
\centering
    \caption{Plots of $\alpha$ vs $\beta$ for $\txbit=1$ obtained at 10 dB and 0 dB SNR with various $\bit$ by solving \eqref{eq:opt_rate_per_tx} for $\Ropt=1.8$ (90\% level). Note that $\ratio$ changes quickly with $\beta$ when $\beta\leq 40$.}
    \label{fig:fixed_Rate_beta_vs_alpha}
\end{figure}

Fig. \ref{fig:fixed_Rate_beta_vs_alpha} shows that to obtain $\Ropt=1.8$ ($90\%$ level for $\txbit=1$), $\ratio$ changes quickly with $\beta$ when $\beta\leq 40$, and slowly otherwise.

\subsubsection{Receiver elements can compensate for $\bit$ and $\snr$}
\begin{figure}
\includegraphics[width=3.5in]{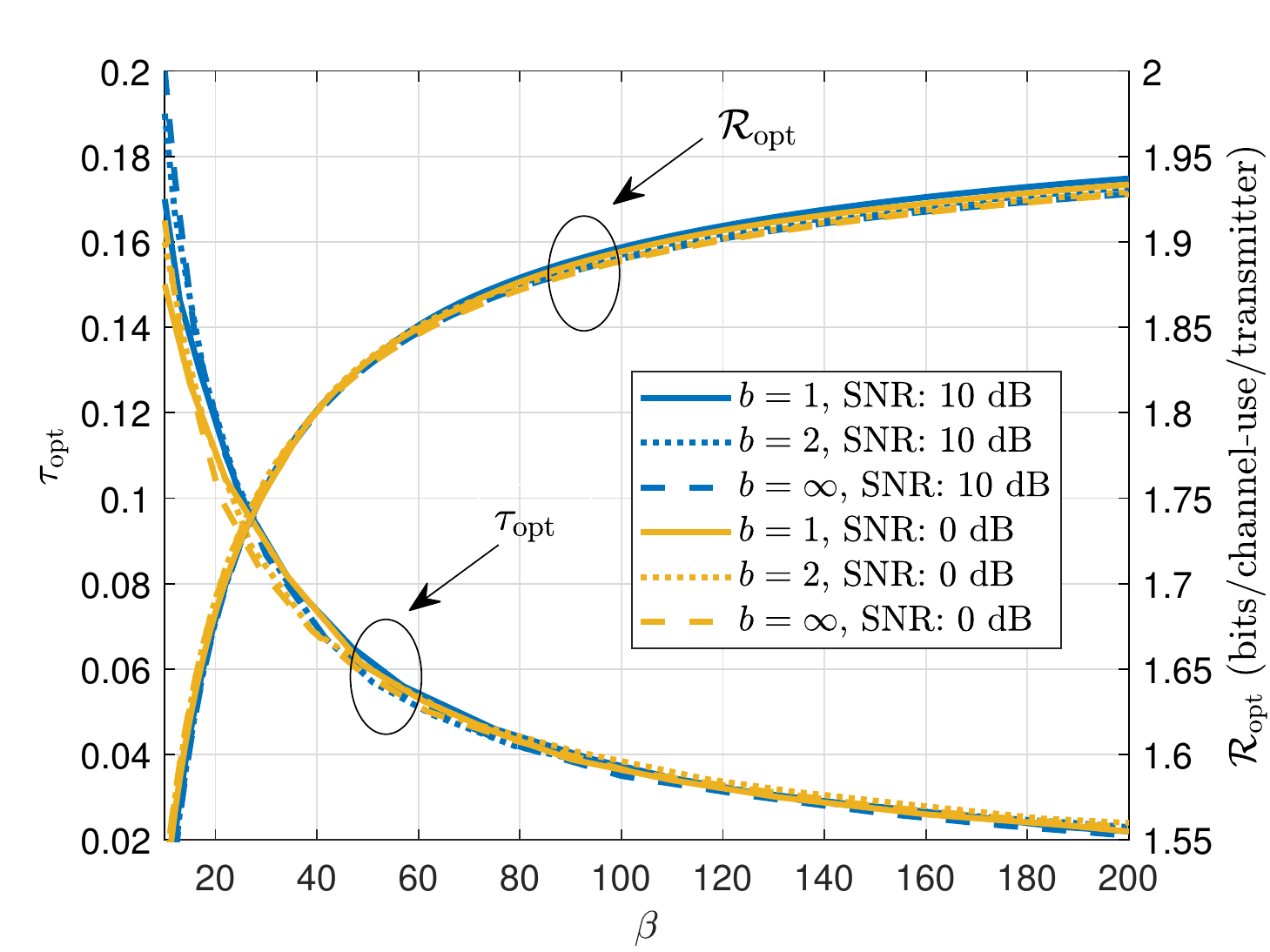}
\centering
    \caption{Plots of $\tauopt$ \eqref{eq:tau_opt_value} vs $\beta$ and $\Ropt$ \eqref{eq:opt_rate_per_tx} vs $\beta$ for $\txbit=1$ at 10 dB and 0 dB SNR with $\bit=1,2,\infty$, where values of $\ratio$ are selected from Fig. \ref{fig:fixed_Rate_beta_vs_alpha} at $\beta=40$. Note that both $\tauopt$ and $\Ropt$ mainly depend on $\beta$ and are not sensitive to $\bit$ and SNR, which indicates that $\ratio$ can be used to compensate for $\bit$ and SNR, independently of $\beta$.}
    \label{fig:fixed_Rate_beta_vs_tau}
\end{figure}

We choose $\ratio$ so that $\Ropt=1.8$ when $\txbit=1$ and $\beta=40$ for a variety of $\bit$ and $\snr$.
Shown in Fig.\ \ref{fig:fixed_Rate_beta_vs_tau} is the corresponding $\tauopt$ and $\Ropt$ as $\beta$ is varied.  Note that the various curves are essentially on top of one another, indicating that the chosen $\alpha$ leads to the same $\tauopt$ and $\Ropt$, independently of $\bit$ and $\snr$.

\subsubsection{Small $\ratio$ is equivalent to a non-fading SISO channel}
\label{subsec:small_ratio_Ropt}
For small $\ratio$, the data rate is limited by the receiver, and we consider the rate per receiver
$(1-\tau)\cIInn{\txA}{\rxA}{}$ 
with unit ``bits/channel-use/receiver", instead of the rate per transmitter 
in \eqref{eq:opt_rate_per_tx}.
The optimal training fraction is still \eqref{eq:tau_opt_value} since $\alpha$ is not a function of $\tau$.  As $\ratio\to 0$,  \eqref{eq:qx_qxtilde_solu_A} yields $\qxtilde\propto\ratio$.
For small $\qxtilde$, we have $\IAWGN(\qxtilde,\ptx(\cdot))= \frac{\qxtilde}{\ln 2}+o(\ratio),$
where $\IAWGN(\qxtilde,\ptx(\cdot))$ is defined in \eqref{eq:MuI_AWGN_def}.
Therefore, \eqref{eq:IXY_value_A} yields
\begin{align}
    \cIInn{\txA}{\rxA}{} \approx \Enta[](0,\varnsequ+\snrequ) - \Enta[]({\snrequ},\varnsequ),
    \label{eq:small_alpha}
\end{align}
where $\snrequ$ and $\varnsequ$ are obtained from \eqref{eq:SNR_var_equivalent}, which does not depend on $\ratio$ or $\ptx(x)$, and $\Enta[](\cdot,\cdot)$ is defined in \eqref{eq:Ent_V_o_def}. The right-hand side of \eqref{eq:small_alpha} is actually the mutual information between the input and output of the following single-input-single-output (SISO) system without any fading: $\rx=f(\sqrt{\snrequ}\tx+\bar{\ns}),$
where $\tx\sim\cC\cN(0,1)$, $\bar{\ns}\sim\cC\cN(0,\varnsequ)$.  We have
\begin{align*}
(1-\tau)\cIInn{\txA}{\rxA}{}&\approx (1-\tau)[\Enta[](0,\varnsequ+\snrequ) - \Enta[]({\snrequ},\varnsequ)],
    \numberthis
    \label{eq:small_alpha_per_rx}
\\
    \tauopt&\approx\argmax_{\tau}(1-\tau)[\Enta[](0,\snrequ+\varnsequ)-\Enta[](\snrequ,\varnsequ)].
    \numberthis
    \label{eq:tauopt_small_ratio}
\end{align*}
Both quantities in \eqref{eq:small_alpha_per_rx} and \eqref{eq:tauopt_small_ratio} are not functions of $\ratio$ or the input distribution $\ptx(x)$.

\begin{figure}
\includegraphics[width=3.5in]{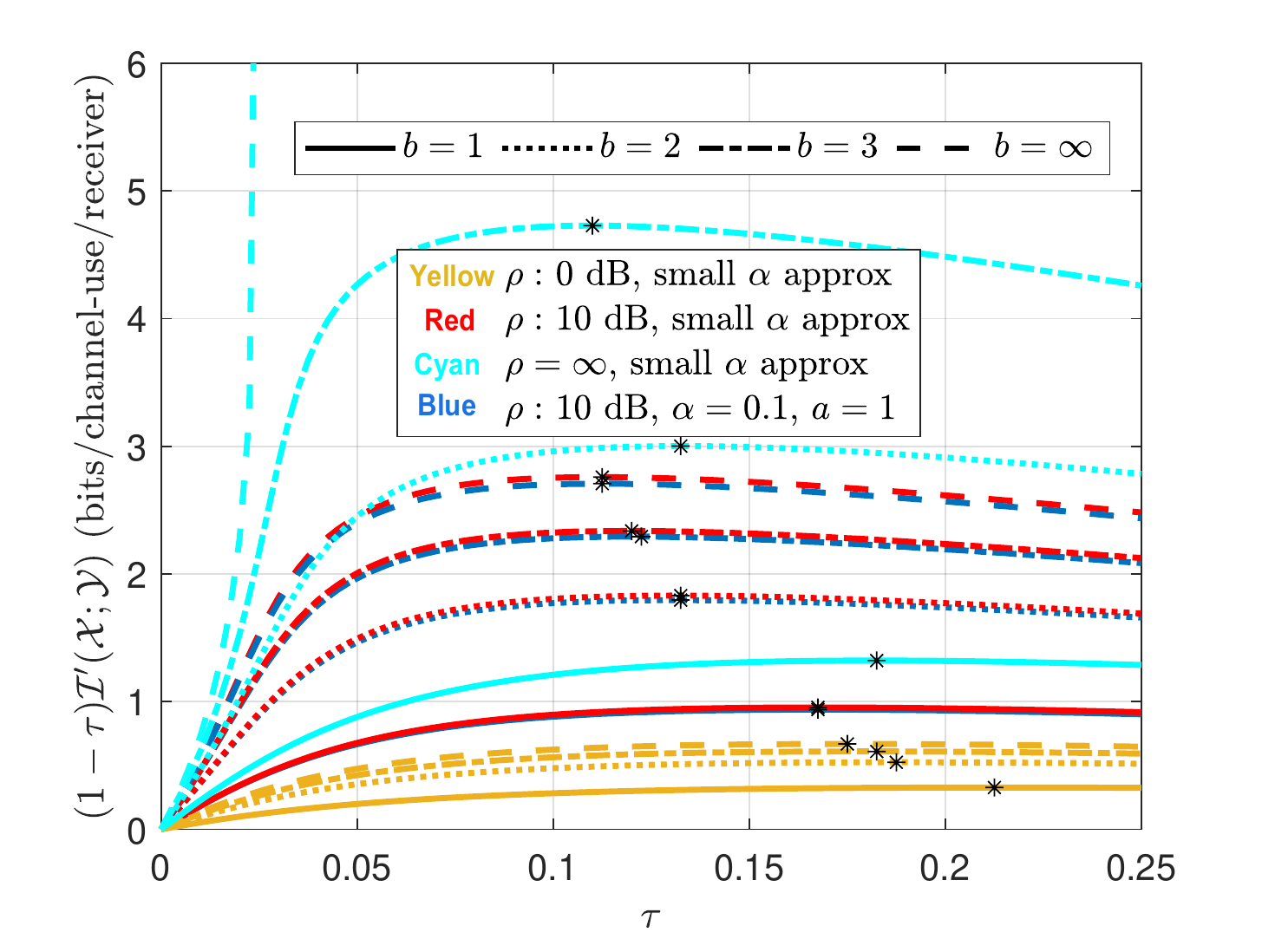}
\centering
    \caption{Small $\ratio$ approximation \eqref{eq:small_alpha_per_rx} of average rate per receiver $(1-\tau)\cIInn{\txA}{\rxA}{}$
    vs $\tau$ for $\rho=0$, 10, $\infty$ dB, for various $\bit$ with $\beta=40$ in yellow, red, and cyan, and the exact value in blue for $\ratio=0.1$ with $\txbit=1$ (QPSK input) at 10 dB SNR.  The values of $\tauopt$ are marked by asterisks. Since \eqref{eq:small_alpha_per_rx} does not depend on $\ratio$ or the input distribution, the value of $\tauopt$ is also insensitive to these quantities. Note that increasing $\bit$ is much more beneficial to rate at high SNR than low SNR. For $\bit<\infty$, the rates per receiver are below $2\bit$ even when $\snr=\infty$ because the quantization effects at the receiver limit our ability to estimate $\chm$.  When $\bit=\infty$ and $\snr=\infty$, perfect channel estimation can be obtained with $\tau=\frac{1}{\beta}=0.025$, and the rate is $\infty$.}
    \label{fig:small_alpha_IXY_vs_tau}
\end{figure}

Fig. \ref{fig:small_alpha_IXY_vs_tau} shows $(1-\tau)\cIInn{\txA}{\rxA}{}$ and its approximations vs $\tau$ with $\beta=40$ for various SNR and $\bit$ for small $\ratio$. Note that the blue and red curves essentially overlap, indicating that the approximation is accurate for $\ratio\leq 0.1$. Note also that the curves are only modestly concave in $\tau$, indicating only mild sensitivity of the rate to the amount of training.

\subsubsection{Optimum training time decreases as $\ratio$ increases}
\label{subsub:tau_vs_alpha}
$\forall\tau>0$, any finite $\txbit$, and any $\bit$, we have
\begin{align*}
    \lim_{\ratio\to\infty}\ratio\cIInn{\txA}{\rxA}{}=2\txbit,
    \numberthis
    \label{eq:large_alpha_rate_saturation}
\end{align*}
which yields $ \lim_{\ratio\to\infty}\Ropt=2\txbit$, and $\lim_{\ratio\to\infty}\tauopt = 0.$
The proof is shown in Appendix \ref{app:large_alpha_rate_and_training}.
In the special case with $\txbit=1$, when $\ratio$ is large, we have
\begin{align}
\text{for } b = \infty, \quad \tauopt&\approx 2\left(\frac{\snr+1}{\snr}\right)^2\frac{\ln\ratio}{\beta\ratio},
\label{eq:tauopt_linear_output}
\\
\text{for } b = 1, \quad
\tauopt&\approx 2\left(\frac{\pi}{2}\frac{\snr+1}{\snr}\right)^2\frac{\ln\ratio}{\beta\ratio}.
\label{eq:tauopt_one_bit_output}
\end{align}
 

\section{Application to Signal Processing}
\label{sec:signal_processing}
We consider an Internet-of-Things (IoT) system \cite{kim2018uplink,liu2018massive,de2018application,uccuncu2017performance}  where wireless devices are used for remote monitoring, and the data captured from those devices can be modeled by \eqref{eq:nonlinear_system_model}, where $\txv_t$ models the transmitted signal at time $t$ from $\tn$ single-element devices, $\rxv_t$ models the received signal from $\rn$ elements at time $t$, and $\chm$ is the unknown wireless channel.  We are specifically interested in the effect of the training time $\Ttone$ on the symbol error rate (SER). \comm{\eqref{eq:Pe_input_training}.}  We define a new quantity
\begin{align}
    \taup={\Tt}/{\tn},
    \label{eq:signal_processing_ratios}
\end{align}
which represents our training time relative to the number of transmitters (sensors). For simplicity, we assume $\varns=1$ throughout this section.

To proceed, we expand the statement of equivalence of Theorem \ref{thm:equivalence_in_Ent} to symbol error probabilities.  We state this equivalence as a conjecture because it is not proven herein, but whose consequences appear to be accurate and useful.
\begin{conject}
\label{conj:equivalent_Pe_x}
For the system \eqref{eq:nonlinear_system_model} with $\Ttone$ input-output training pairs $(\txm_{\Ttone},\rxm_{\Ttone})$, we have
\begin{align}
 \lim_{\Tlim\to\infty}P_{e,x}  = \lim_{\Tlim\to\infty}\bar{P}_{e,x},
 \numberthis
 \label{eq:equ_in_SER}
\end{align}
where $P_{e,x}$ and $\bar{P}_{e,x}$ are average probability of errors defined as
\begin{align*}
    P_{e,x}&=\frac{1}{\tn}\sum_{m=1}^{\tn}\E P(\tx_{\Ttone+1,m}\neq\hat{\tx}_{\Ttone+1,m}|\txm_{\Ttone},\rxm_{\Ttone},\rxv_{\Ttone+1}), \ 
    \bar{P}_{e,x}=\frac{1}{\tn}\sum_{m=1}^{\tn}\E P(\txequ_{m}\neq {\hatbar{\tx}}_m|\chm,\rxvequ),
    \numberthis
    \label{eq:Pe_input_training&equivalent_SER_def}
\end{align*}
$\tx_{\Ttone+1,m}$ and $\txequ_{m}$ are the $m$th elements of $\txv_{\Ttone+1}$ and $\txvequ$ in the equivalent system \eqref{eq:equivalent_system_thm}, $\hat{\tx}_{\Ttone+1,m}$ and ${\hatbar{\tx}}_m$ are defined as
\begin{align*}
    \hat{\tx}_{\Ttone+1,m}&=\argmax_{x}P(\tx_{ \Ttone+1,m}=x|\txm_{\Ttone},\rxm_{\Ttone},\rxv_{\Ttone+1}), \quad
    {\hatbar{\tx}}_m=\argmax_{x}P(\txequ_{m}=x|\chm,\rxvequ).
    \numberthis
    \label{eq:MPM_detection_rule&MPM_detection_rule_equivalent}
\end{align*}
\end{conject}
The quantities $\hat{\tx}_{\Ttone+1,m}$ and ${\hatbar{\tx}}_m$ in \eqref{eq:MPM_detection_rule&MPM_detection_rule_equivalent} are called the marginal posterior mode (MPM) detectors of $\tx_{\Ttone+1,m}$ and ${\bar{\tx}}_m$ and minimize the SER $P_{e,x}$ and $\bar{P}_{e,x}$ in \eqref{eq:Pe_input_training&equivalent_SER_def}; see \cite{tanaka2001analysis,tanaka2002statistical}.
Equation \eqref{eq:equ_in_SER} conjectures the equivalence between unknown-$\chm$ and known-$\chm$ probabilities of error in estimating the input vector, conditioned on the output vector.  The probability of error $\bar{P}_{e,x}$ in \eqref{eq:Pe_input_training&equivalent_SER_def} is calculated using the transformations of $\snr$ and $\varns$ to $\snrequ$ and $\varnsequ$ as stipulated in Theorem \ref{thm:equivalence_in_Ent}. 
As is true for Theorem \ref{thm:equivalence_in_Ent}, the value of Conjecture \ref{conj:equivalent_Pe_x} is its ability to convert the analysis of a system with unknown $\chm$ to a (presumably simpler) system with known $\chm$.

\subsection{SER analysis using Conjecture \ref{conj:equivalent_Pe_x}}
The SER of a large-scale system \eqref{eq:nonlinear_system_model} with known $\chm$ is analyzed in \cite{wen2016bayes}.  We may leverage these results by using the equivalence \eqref{eq:equ_in_SER} to convert the SER of a system with unknown $\chm$ to the SER of its equivalent system. The equivalent system is defined in \eqref{eq:equivalent_system_thm}, where $\snrequ$ and $\varnsequ$ can be obtained from Steps 1)-2) in Section~\ref{sec:step_by_step_computation} with $\tau\beta$ replaced by $\taup$ defined in \eqref{eq:signal_processing_ratios}.  Then, according to \cite{wen2016bayes}, $\bar{P}_{e,x}$ defined in \eqref{eq:Pe_input_training&equivalent_SER_def} is obtained by analyzing $ \rx = \sqrt{\qxtilde}\tx+\ns,$
where $\qxtilde$ is found from \eqref{eq:qx_qxtilde_solu_A}, the distribution of $\tx$ is the same as that of elements of $\txvequ$, and $\ns\sim\cC\cN(0,1)$.  

For $\txbit=1$ (QPSK modulation), we have
\begin{align}
    \lim_{\Tlim\to\infty} \bar{P}_{e,x} = 2Q(\sqrt{\qxtilde}) - [Q(\sqrt{\qxtilde})]^2,
    \label{eq:SER_QPSK_training}
\end{align}
where $Q(\cdot)$ is the tail distribution function of the standard Gaussian.  Then \eqref{eq:equ_in_SER} yields that $\lim\limits_{\Tlim\to\infty} P_{e,x}$ is also given by \eqref{eq:SER_QPSK_training}.

\begin{figure}
\includegraphics[width=3.5in]{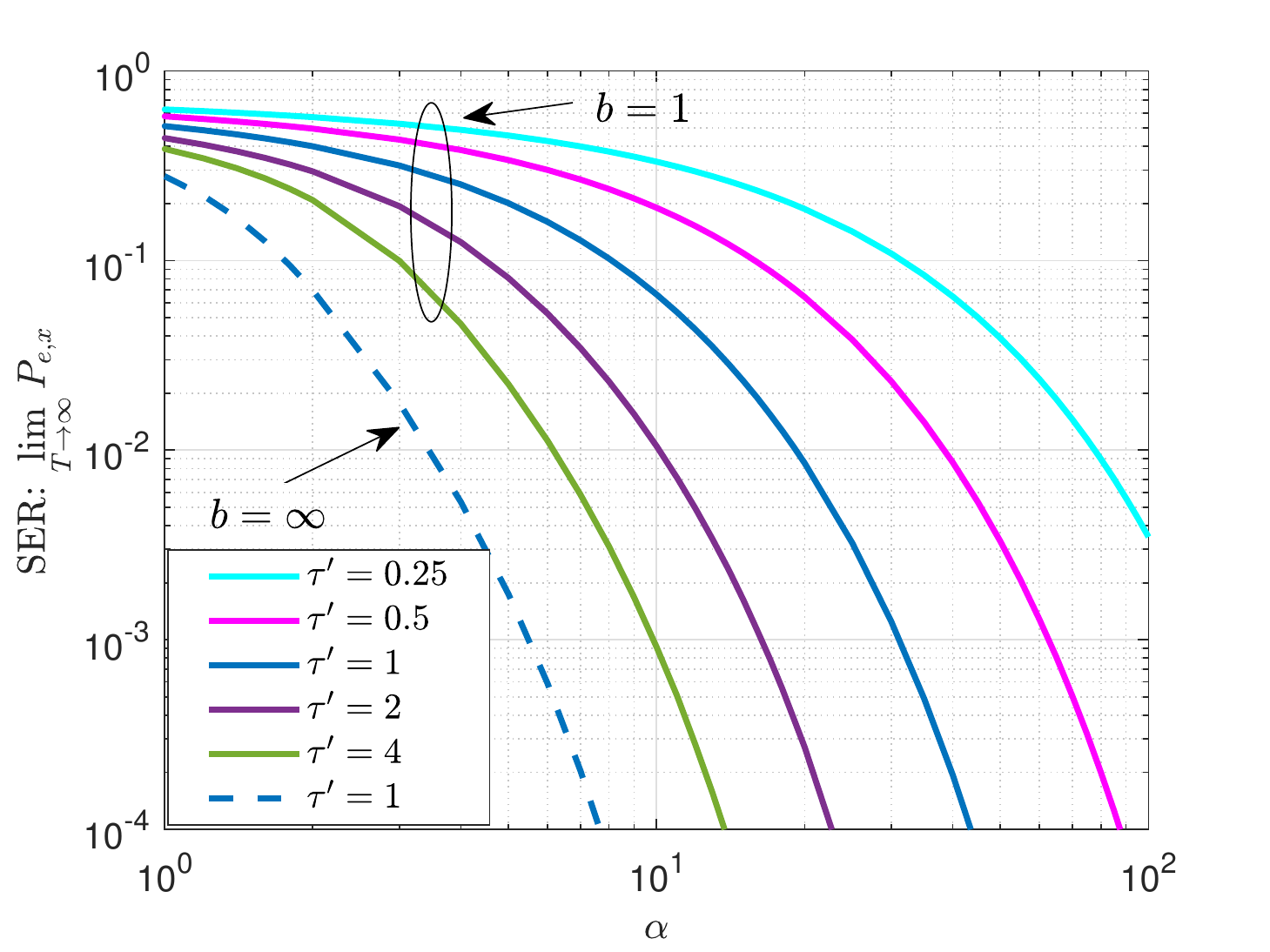}
\centering
    \caption{Plots of SER in a large-scale limit $\lim\limits_{\Tlim\to\infty} P_{e,x}$ \eqref{eq:SER_QPSK_training} vs $\ratio$ with $\txbit=1$ for $\bit=1, \infty$ and $\taup=0.25, 0.5, 1, 2, 4$ at 10 dB SNR. Note that the SER can be arbitrarily small as long as $\ratio$ is large enough, as is indicated in \eqref{eq:large_ration_SER_approx}, even for $\bit=1$ and $\taup<1$ where the number of training signals is smaller than the number of transmitters.}
    \label{fig:alpha_vs_SER_QPSK}
\end{figure}

When $\ratio$ is large, \eqref{eq:qx_qxtilde_solu_A} yields that $\msea{\tx}$ decays exponentially to zero for $\txbit=1$, and $\qxtilde\approx\ratio\snrequ\cdot\chi(\snrequ,\varnsequ).$
Then, SER in \eqref{eq:SER_QPSK_training} can be approximated as
\begin{align}
    \lim_{\Tlim\to\infty} P_{e,x} \approx 2Q(\sqrt{\ratio\snrequ\cdot\chi(\snrequ,\varnsequ)}),
    \label{eq:large_ration_SER_approx}
\end{align}
where $\snrequ$ and $\varnsequ$ obtained from \eqref{eq:SNR_var_equivalent} are not functions of $\ratio$, and $\chi(\cdot,\cdot)$ is defined in \eqref{eq:chi_value} for finite $\bit$ and in \eqref{eq:linear_Ent_chi} for $\bit=\infty$.  This shows that the SER can be made arbitrarily small when $\ratio$ increases, no matter how small $\bit$ and $\taup$ are.  It is perhaps surprising that all transmitted symbols can be correctly identified even if the number of training signals is smaller than the number of transmitters ($\taup<1$).
Fig. \ref{fig:alpha_vs_SER_QPSK} shows SER vs $\ratio$ with $\txbit=1$ for $\bit=1, \infty$ and $\taup=0.25, 0.5, 1, 2, 4$, where we can see that the SER can be arbitrarily small as long as $\ratio$ is large enough.

\begin{figure}
\includegraphics[width=3.5in]{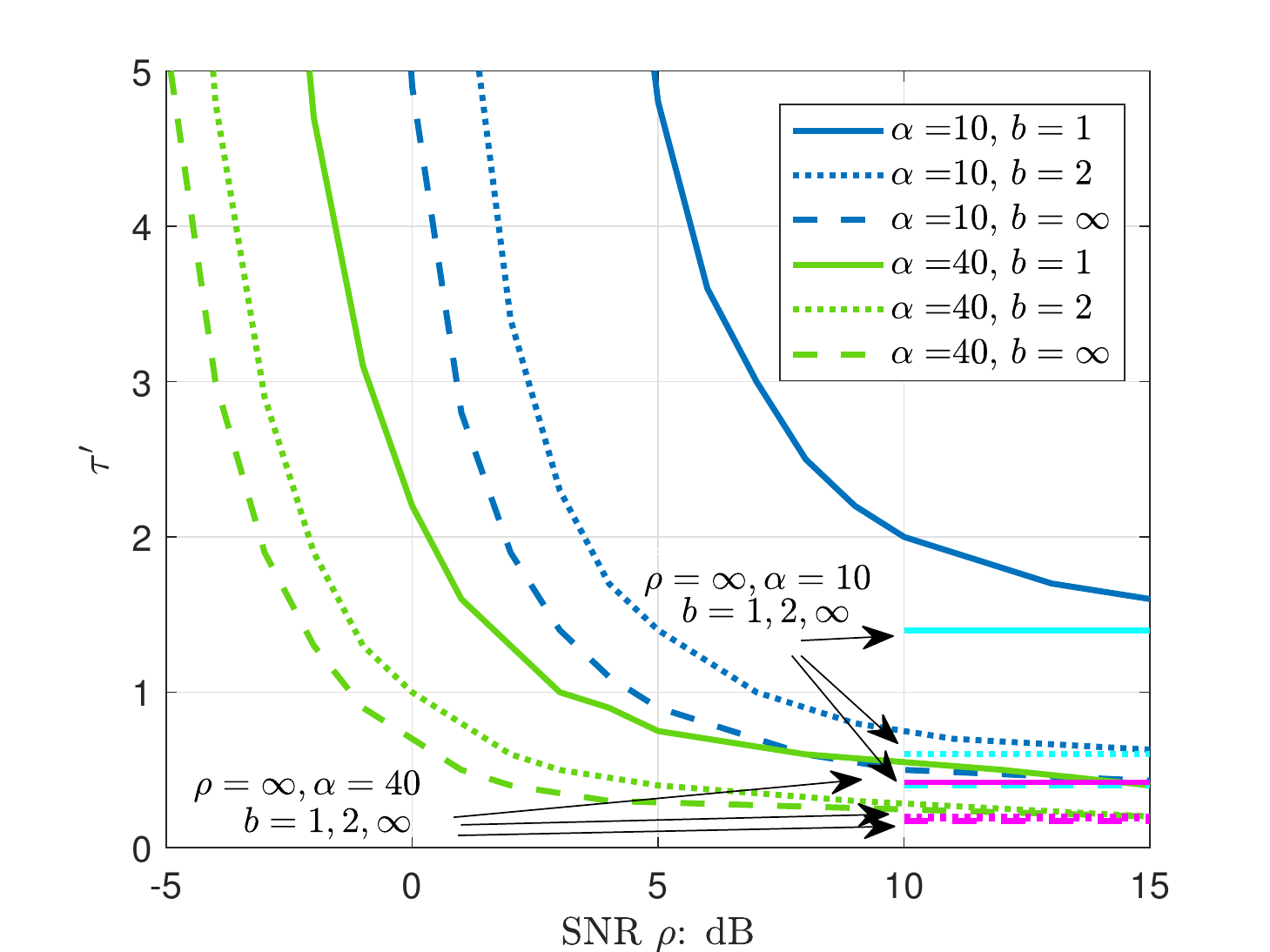}
\centering
    \caption{Plots of $\taup$ vs SNR for $\txbit=1$ with $\bit=1,2,\infty$ and $\ratio=10, 40$ obtained by solving \eqref{eq:qh_qhtilde_solu_A&mse_G_vs_qh}--\eqref{eq:qx_qxtilde_solu_A} together with \eqref{eq:SER_QPSK_training} for  $\lim\limits_{\Tlim\to\infty} P_{e,x}=0.01$, where $\tau\beta$ in \eqref{eq:qh_qhtilde_solu_A&mse_G_vs_qh} is replaced by $\taup$. Note that $\taup$ increases dramatically as SNR decreases at low SNR and $\taup$ does not change much with SNR at high SNR, with asymptotes shown in cyan for $\ratio=10$ and in magenta for $\ratio=40$ when $\snr=\infty$. Therefore, to obtain 1\% SER with short training, the system should be operated at SNR above the ``knee", which can be considered as the SNR that achieves 1\% SER with $\taup=2$, called the ``critical SNR". The critical SNR can be reduced by either increasing $\bit$ or $\ratio$. Increasing $\bit$ can only reduce the critical SNR by a finite amount, while increasing $\ratio$ can make the critical SNR arbitrarily small.}
    \label{fig:SNR_vs_tau}
\end{figure}

The required $\taup$ for training vs SNR to achieve 1\% SER with $\txbit=1$ for $\bit=1, 2, \infty$ and $\ratio=10, 40$ is shown in Fig. \ref{fig:SNR_vs_tau}.    Observe that $\taup$ is very large at low SNR and the SNR at which 1\% SER can be achieved with $\taup=2$ is considered as the ``critical SNR", below which $\taup$ increases dramatically as SNR decreases to maintain the SER. The critical SNR can be reduced by increasing $\bit$ or $\ratio$. By increasing $\bit$, the critical SNR can only be reduced by a limited value, while by increasing $\ratio$, the critical SNR can be arbitrarily small.

\begin{figure}
\includegraphics[width=3.5in]{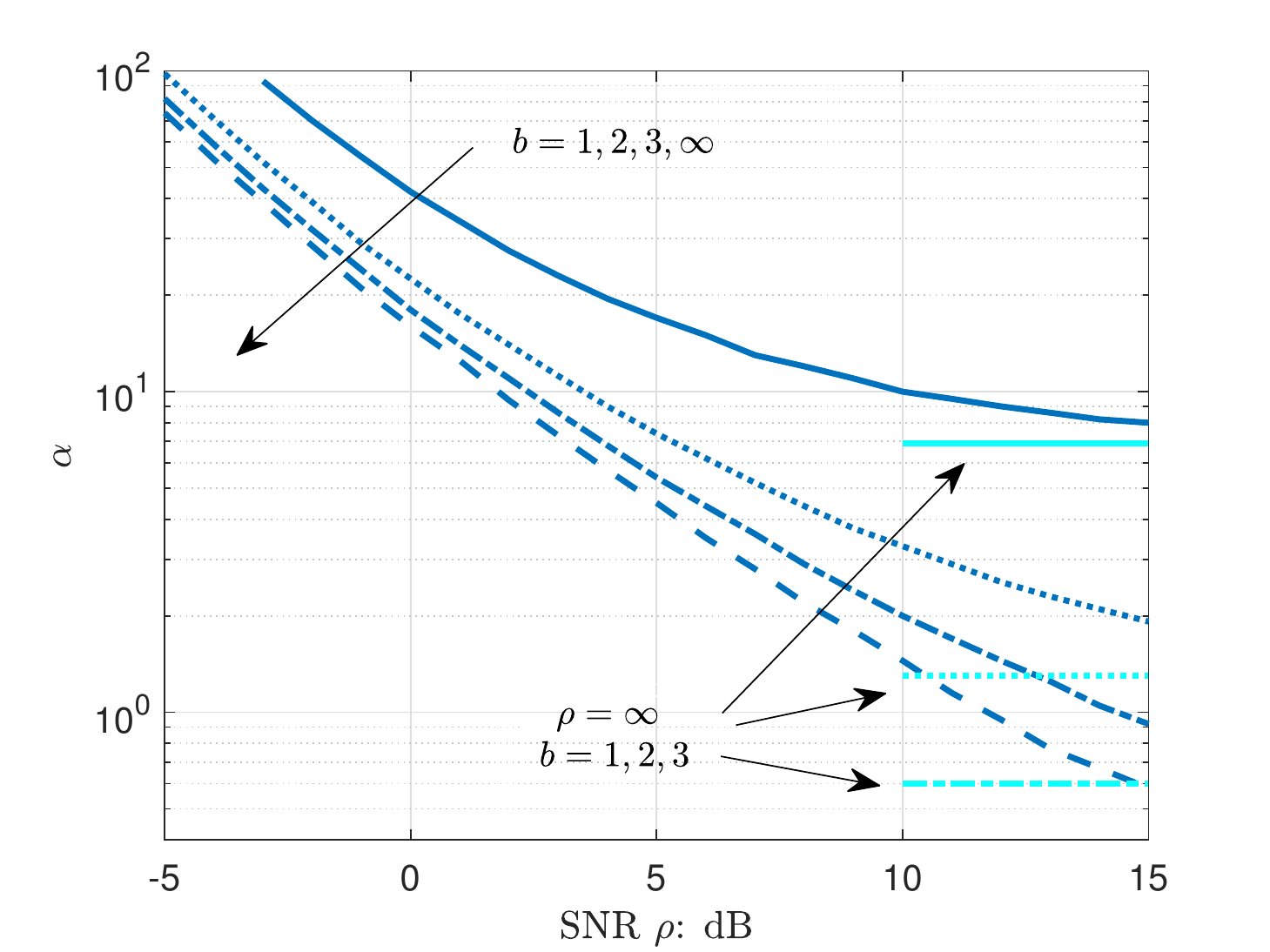}
\centering
    \caption{Plots of $\ratio$ vs SNR with $\txbit=1$ and $\taup=2$ that achieves 1\% SER \eqref{eq:SER_QPSK_training}. Note that $\ratio$ decreases as SNR increases and reaches asymptotes shown in cyan as $\snr=\infty$ with $\bit=1,2,$ and $3$, and the asymptotes are the result of the quantization noise at the receiver, similar to that shown in Fig. \eqref{fig:fixed_tau_opt_SNR_vs_alpha_2}. Perfect channel estimation can be obtained for linear receivers $(b=\infty)$, and therefore there is no asymptote.}
    \label{fig:alpha_vs_SNR_dB}
\end{figure}

The required $\ratio$ vs SNR to obtain 1\% SER with $\txbit=1$ and $\taup=2$ for various $\bit$ is shown in Fig. \ref{fig:alpha_vs_SNR_dB}. It is shown that $\ratio$ decreases as SNR increases for all $\bit$, and at high SNR, the values of $\ratio$ decrease faster for larger $\bit$. 
In the extreme case when the thermal noise is negligible $(\snr=\infty)$, the quantization noise at the receivers prohibits the values of $\ratio$ from going to zero for $\bit=1,2,3$, and $\ratio$ goes to 0 for $\bit=\infty$ (linear receivers), where the channel can be estimated perfectly.


\subsection{Evidence of accuracy of the conjecture}
\label{sec:finite_OK}

For $\txbit=1$, we show numerically that $\lim\limits_{\Tlim\to\infty} \bar{P}_{e,x}$ obtained from \eqref{eq:SER_QPSK_training} is accurate even for reasonable values of $\tn$.  We need to compute the estimate $\hat{\tx}_{\Ttone+1,m}$ shown in \eqref{eq:MPM_detection_rule&MPM_detection_rule_equivalent}, but this is complicated  even for small $\tn$.  We instead use an approximation: first, we obtain a channel estimate $\tilde{\chm}$ by using the  transmitted and received training signals $(\txm_{\Ttone},\rxm_{\Ttone})$; then, we treat the estimated channel $\tilde{\chm}$ as the true channel, while the channel estimation error is treated as part of the additive Gaussian noise; finally, we estimate each element of the transmitted data vector $\txv_{\Ttone+1}$ by using $\tilde{\chm}$ and the corresponding received vector $\rxv_{\Ttone+1}$ through the following model:
\begin{align}
    \rxv_{\Ttone+1} = \bff\left(\sqrt{{\snr}/{\tn}}\tilde{\chm}\txv_{\Ttone+1}+\tilde{\nv}_{\Ttone+1}\right),
    \label{eq:equivalent_MMSE_detection_model}
\end{align}
where $\tilde{\nv}_{\Ttone+1}$ includes additive noise $\nv_{\Ttone+1}$ and the channel estimation error. 

We apply a generalized approximate message passing (GAMP)-based algorithm proposed in \cite{wen2016bayes} twice, once to obtain $\tilde{\chm}$, and then again for the estimate of $\txv_{\Ttone+1}$ from \eqref{eq:equivalent_MMSE_detection_model}. This algorithm as used in \cite{wen2016bayes} is applied to joint channel and data estimation by processing the received training and data signals jointly. We use it to estimate the channel from only the training signals, and estimate the data from only the received data signals (treating the estimated channel as known). The three steps of our algorithm are summarized here: first, we obtain $\tilde{\chm}$ from the training signals $(\txm_{\Ttone},\rxm_{\Ttone})$ by using the GAMP-based algorithm; second, we obtain $\msea{\chm}$ by solving \eqref{eq:qh_qhtilde_solu_A&mse_G_vs_qh} with $\tau\beta$ replaced by $\taup=\frac{\Ttone}{\tn}$, and we then model the relationship between $\txv_{\Ttone+1}$ and $\rxv_{\Ttone+1}$ as \eqref{eq:equivalent_MMSE_detection_model}; third, we estimate elements of $\txv_{\Ttone+1}$ from $\rxv_{\Ttone+1}$ by applying GAMP to the model \eqref{eq:equivalent_MMSE_detection_model}, followed by an element-wise hard-decision. Note that we treat the variances of elements of $\tilde{\chm}$ and $\tilde{\nv}_{\Ttone+1}$ as $(1-\msea{\chm})$ and $(1+\snr\msea{\chm})$ while using GAMP in the third step, where the values of the variances are obtained from Conjecture \ref{conj:equivalent_Pe_x}.  
Since our algorithm applies GAMP twice, we simply call it GAMP2.

\begin{figure}
\includegraphics[width=3.5in]{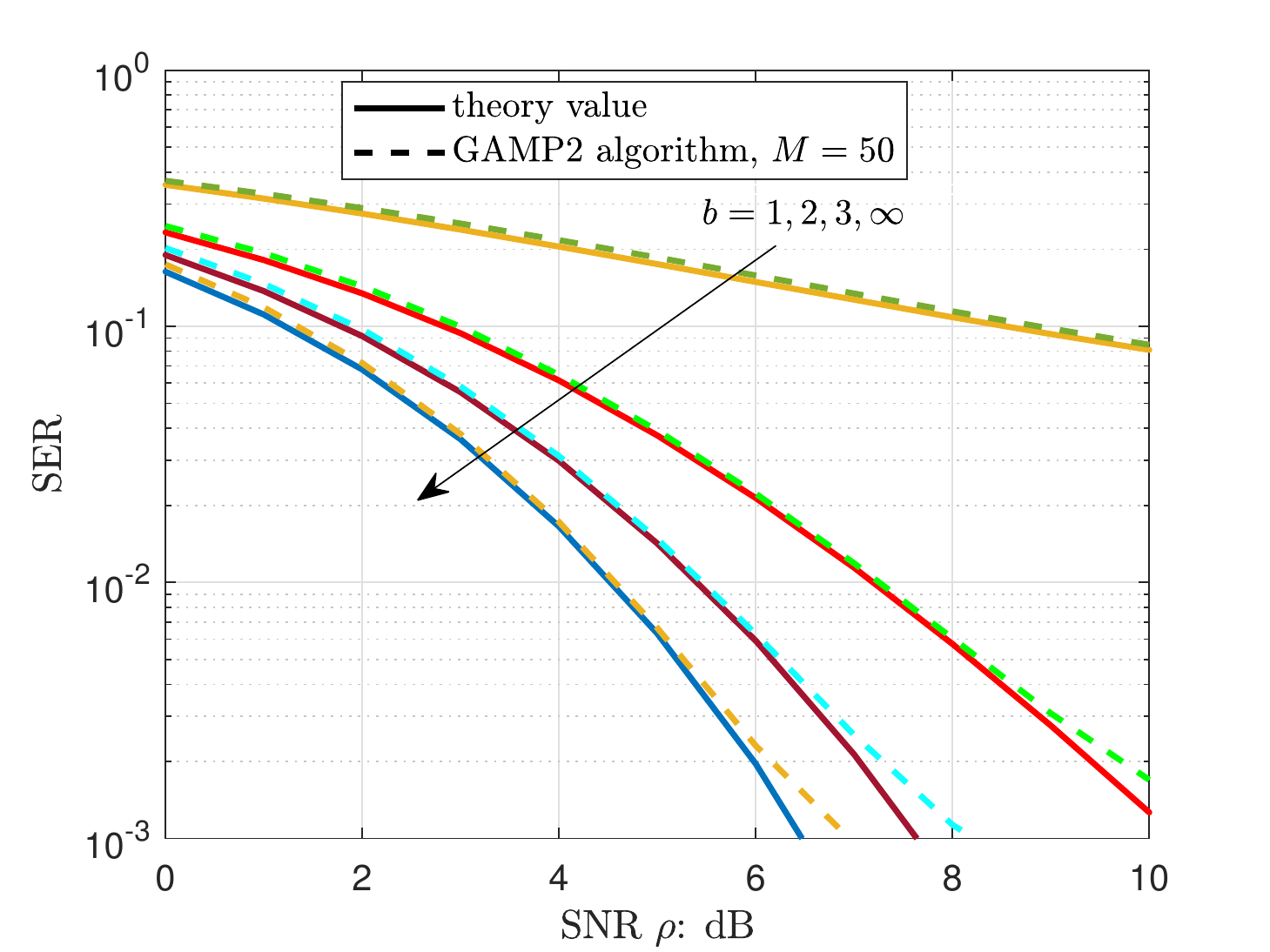}
\centering
    \caption{Plots of SER vs SNR with $\taup=2,\ratio=5$ and $\txbit=1$ for $\bit=1,2,3,\infty$ obtained from the theory value \eqref{eq:SER_QPSK_training} (solid lines) and from the GAMP2 algorithm with $\tn=50$ (dashed lines). The performance of GAMP2 algorithm is very close to the theory analysis with accuracy of SER up to about 0.3\%.}
    \label{fig:GAMP_vs_theory}
\end{figure}

The numerical results of SER vs SNR with 
 $\taup=2,\ratio=5$ and $\txbit=1$ (QPSK modulation) for $\bit=1,2,3,\infty$ obtained from \eqref{eq:SER_QPSK_training} in a large-scale limit theory analysis and obtained from the GAMP2 algorithm with $\tn=50$ ($\rn=\ratio\tn=250, \Ttone=\taup\tn=100$) are shown in Fig. \ref{fig:GAMP_vs_theory}. The numerical values of the SER for the GAMP2 algorithm are computed by averaging the SER obtained from GAMP2 algorithm for each realization of $(\chm, \txm_{\Ttone}, \rxm_{\Ttone}, \txv_{\Ttone+1}, \rxv_{\Ttone+1})$ over $10^6$ realizations.  The GAMP2 results and theoretical analysis in \eqref{eq:SER_QPSK_training} clearly track each other closely.

\section{Discussions and Conclusions}

We leveraged the results in \cite{gaopart1} and \cite{wen2016bayes} to derive a variety of results on training in communication and signal processing models with quantization at the input and output of the system.  Our results used an equivalence relationship between an unknown-channel-with-training model and a known-channel model that applies to large-scale systems.

We have conjectured the equivalence shown in Theorem  \ref{thm:equivalence_in_Ent} can be generalized to other macroscopic quantities such as probability of error and used this conjecture in a signal processing application. Evidence of the accuracy of the conjecture was provided.  It would be of interest to see whether this conjecture can be proved.

We believe that the equivalence shown in Theorem \ref{thm:equivalence_in_Ent} for the model \eqref{eq:nonlinear_system_model} can be generalized beyond quantizers to other nonlinear functions $f(\cdot)$.  In particular, since a quantizer with sufficiently high resolution and number of levels can be used to approximate a well-behaved monotonic function, it is conceivable that the theorem can readily be adapted to any monotonic function. We view this as a possible avenue for future work.

\appendices

\section{Proof of Theorem \ref{thm:equivalence_in_Ent}}
\label{app:proof_equivalence}


We prove \eqref{eq:equi_in_MuI}--\eqref{eq:equi_in_Ent_Y} by deriving the expressions of both sides of the equality and noticing that they are the same.

\subsection{Definition of some functions}
\label{subsec:function_def}
To compute the quantities in \eqref{eq:equi_in_MuI}--\eqref{eq:equi_in_Ent_Y}, we first define $\IAWGN(\qtilde,p(\cdot))$, $\mse(\qtilde,p(\cdot))$, $\Enta[](\gainv,s),$ and $\chi(\gainv,s)$ as below.  


Let $\IAWGN(\qtilde,\ptx(\cdot))$ and $\mse(\qtilde,\ptx(\cdot))$ be
\begin{align}
    \IAWGN(\qtilde,\ptx(\cdot))&=-\E_{\rx}[\log_2\E_x(e^{-|y-\sqrt{\qtilde}x|^2})]-\log_2e,
    \label{eq:MuI_AWGN_def}
\\
    \mse(\qtilde,\ptx(\cdot))&=\E_{x,y}(|x-\int x\cdot p(x|y)dx|^2),
    \numberthis
    \label{eq:MSE_AWGN_def}
\end{align}
where $x$ is a complex number whose real and imaginary parts $x_{\rm R}$ and $x_{\rm I}$ are \iid with distribution $\ptx(\cdot)$.  The distribution of $x$ is
$p(x)=\ptx(x_{\rm R})\cdot \ptx(x_{\rm I})$,
and the joint distribution of $(x,y)$ is
\begin{align}
    p(x,y)=p(x)\cdot \frac{1}{\pi}e^{-|\rx-\sqrt{\qtilde}x|^2},
    \label{eq:p_wy_joint}
\end{align}
and $p(x|y)$ is the conditional distribution of $x$ conditioned on $y$.

Note that $(x,y)$ that satisfies the joint distribution \eqref{eq:p_wy_joint} can be modeled as 
\begin{align}
    \rx=\sqrt{\qtilde}x+\ns,
    \label{eq:AWGN_tilde_q1}
\end{align}
where $\ns\sim\cC\cN(0,1)$ is independent of $x$. \eqref{eq:AWGN_tilde_q1} describes a single-input-single-output (SISO) additive white Gaussian noise (AWGN) channel. $\IAWGN(\qtilde,\ptx(\cdot))$ shows the mutual information between the $x$ and $\rx$, and 
$\mse(\qtilde,\ptx(\cdot))$ shows the mean-square error (MSE) of the MMSE estimate of $\tx$ conditioned on $\rx$.
$\IAWGN(\qtilde,\ptg(\cdot))$ and $\mse(\qtilde,\ptg(\cdot))$ are defined similarly by replacing $\ptx(\cdot)$ with $\ptg(\cdot)$.


For a $\bit$-bit uniform quantizer $f(\cdot)$, we define $\Enta[](\gainv,s)$ and $\chi(\gainv,s)$ as 
\begin{align*}
    \Enta[](\gainv,s)&=-2\sum_{k=1}^{2^{\bit}}\int_{\cR}dz\frac{e^{-\frac{z^2}{2}}}{\sqrt{2\pi}}\Psi_k(\sqrt{\gainv}z,s) \log_2\Psi_k(\sqrt{\gainv}z,s),
    \numberthis
    \label{eq:Ent_V_o_def}\\
        \chi(\gainv,s)&=\sum_{k=1}^{2^{\bit}}\int_{\cR}dz\frac{e^{-\frac{z^2}{2}}}{\sqrt{2\pi}}\frac{(\Psi'_k(\sqrt{\gainv} z,s))^2}{\Psi_k(\sqrt{\gainv} z,s)},
        \numberthis
    \label{eq:chi_value}
\end{align*}
where 
\begin{align*}
    \Psi_k(w,s)&=\Phi\left(\frac{\sqrt{2}r_k-w}{\sqrt{s}}\right)-\Phi\left(\frac{\sqrt{2}r_{k-1}-w}{\sqrt{s}}\right),
    \numberthis
    \label{eq:psi_func} \\
    \Psi'_k(w,s)&=\frac{e^{-\frac{(\sqrt{2}r_k-w)^2}{2s}}-e^{-\frac{(\sqrt{2}r_{k-1}-w)^2}{2s}}}{\sqrt{{2\pi s}}},
    \numberthis
    \label{eq:psi_prime_b}
\end{align*}
with $r_k$ defined in \eqref{eq:threshold_levels}, 
 and $\Phi(\cdot)$ is the cumulative distribution function (CDF) of $\cN(0,1)$.

For a linear function $f(w)=w$ ($\bit=\infty$), $\Enta[](\gainv,s)$ and $\chi(\gainv,s)$ are defined as
\begin{align}
    \Enta[](\gainv,s)=\log_2 (\pi es),\quad \chi(\gainv,s)= {1}/{s},
    \label{eq:linear_Ent_chi}
\end{align}
and the  computed $\EntInnC{\txA}{\rxA}{}{}$ and $\EntInnC{\txA}{\rxA}{^+}{}$ are then differential entropies. 



\subsection{Left sides of \eqref{eq:equi_in_MuI}--\eqref{eq:equi_in_Ent_Y}}
We first derive expressions of $\cIInn{\txA}{\rxA}{}, \EntInnC{\txA}{\rxA}{ ^+}{},$ and $\EntInnC{\txA}{\rxA}{}{}$ by using theorems developed in \cite{gaopart1}, which can be computed from a single entropy $\EntR(\rxA_{\dy}|\txA)=\EntR(\rxA_\dy|\txA_{\dx=1})$, where $\EntR(\rxA_\dy|\txA_\dx)=\lim_{\Tlim\to\infty}\frac{1}{\rn\Tt}\Ent(\rxm_{\ceil{\dy\Tt}}|\txm_{\ceil{\dx\Tt}}).$
$\EntR(\rxA_{\dy}|\txA)$ can be obtained from a quantity called asymptotic free entropy $\cF$ through
\begin{align}
    \EntR(\rxA_\dy|\txA) = -\frac{\cF(\tau\beta,(\dy-1)\tau\beta)}{\ratio\tau\beta},
    \label{eq:Ent_Y_cond_X_tau}
\end{align}
where $\cF(\betat,\betad)$ is defined as $\cF(\betat,\betad) = \lim_{\tn\to\infty}\frac{1}{\tn^2}\E\log_2 p(\rxm_{\ceil{(\betat+\betad)\tn}}|\txm_{\ceil{\betat\tn}}),$
 which has been computed as (44) in \cite{wen2016bayes}, and is continuous in $\betat$ and $\betad$.

Following the steps in \cite[Section~III]{gaopart1}:

\begin{itemize}
    \item [1)]

From the model \eqref{eq:nonlinear_system_model}, it is clear that Assumption A1 in \cite{gaopart1} is met, i.e.
\begin{align*}
 &p(\rxm_{\Tb}|\txm_{\Tb};\chm)=\prod_{t=1}^{\Tb}p(\rxv_t|\txv_t;\chm),\quad p(\txm_\Tb)=p(\txm_{\Tt})\prod_{t=\Tt+1}^{\Tb}p(\txv_t),
\end{align*}
where $p(\rxv_t|\txv_t;\chm)$ is a fixed conditional distribution for all $t=1,2,\ldots,\Tb$, and $p(\txv_t)$ is a fixed distribution for all $t=\Tt+1,\Tt+2,\ldots,\Tb$. 
The dimension of $\chm$ depends on the blocklength $\Tb$ through \eqref{eq:ratios}. 
Furthermore, the input $\txv_t$ are \iid for all $t$. Then, we can express the system by a set of distributions defined as
\begin{align}
    \cP(\Tb,\tau)=\{p(\rxv|\txv;\chm),p(\chm),p(\txv)\},
    \label{eq:P_set_iid_A_input}
\end{align}
which are the conditional distribution of the system, the distribution of $\chm$, and the input distribution. 

\item[2)]
For given $\ratio$ and $\beta$, we define $F(\tau,\dy)=\EntR(\rxA_\dy|\txA),$
where the entropy $\EntR(\rxA_\dy|\txA)$ is computed through  $\cP(\Tb,\tau)$ defined in \eqref{eq:P_set_iid_A_input}. Then, \eqref{eq:Ent_Y_cond_X_tau} yields
\begin{align}
    F(\tau,\dy)=-\frac{\cF(\tau\beta,(\dy-1)\tau\beta)}{\ratio\tau\beta},
    \label{eq:F_tau_P}
\end{align}
\item[3)]
Theorem 2 in \cite{gaopart1} then yields $ \EntR(\rxA_\dy|\txA_{\dx})=u\cdot F\left(u\tau,\frac{\dy-u}{\dx}+1\right)$
for all $\dy,\dx>0$, where $u=\min(\dx,\dy)$.
Then, \eqref{eq:F_tau_P} yields $ \EntR(\rxA_\dy|\txA_{\dx})=
      -\frac{1}{\ratio\tau\beta}\cF(u\tau\beta,(\dy-u)\tau\beta).$
\item[4)]
Corollary 1(b) in \cite{gaopart1} then yields
\begin{align*}
    \EntInnC{\txA}{\rxA}{\dx}{\dy}=\begin{cases} 
      \frac{-1}{\ratio\tau\beta}\frac{\partial}{\partial \dy}\cF(\dy\tau\beta,0), &\dy< \dx; \\
      \frac{-1}{\ratio\tau\beta}\frac{\partial}{\partial \dy}\cF(\dx\tau\beta,(\dy-\dx)\tau\beta), & \dy> \dx.
      \end{cases}
\end{align*}
for all $\dy,\dx>0$. 
Since $\txv_t$ is independent of $(\txv_k,\rxv_k)$ when $t\neq k$, we have $\EntInnC{\txA}{\rxA}{\dx}{\dy}=\EntInnC{\txA}{\rxA}{\dy^+}{\dy},$
for all $\dx>\dy>0$.
\item[5)]
Using $\cF(\cdot,\cdot)$ in \cite{wen2016bayes} and Corollary 1(b) in \cite{gaopart1}, we can verify that 
 \begin{align*}
 \EntInnC{\txA}{\rxA}{^+}{}  = \lim_{\dy\downto 1} \EntInnC{\txA}{\rxA}{\dy^+}{\dy},
\end{align*}
and
\begin{align}
    \EntInnC{\txA}{\rxA}{^+}{} = -\frac{1}{\ratio\tau\beta}\frac{\partial}{\partial \dy}\cF(\dy\tau\beta,0)|_{\dy=1}.
    \label{eq:Ent_Y_cond_X}
\end{align}
Moreover, using $\EntInnC{\txA}{\rxA}{\dx}{\dy}$, we have 
\begin{align*}
    \lim_{\dy\downto 1}\EntInnC{\txA}{\rxA}{}{\dy}=\lim_{\dx\upto 1}\EntInnC{\txA}{\rxA}{\dx}{}.
\end{align*}
Thus, Assumption A2 is met via Corollary 1(c) in \cite{gaopart1}, i.e.
\begin{align*}
  \text{A2:}
  \qquad\quad \EntInnC{\txA}{\rxA}{^+}{}  &= \lim_{\dy\downto 1} \EntInnC{\txA}{\rxA}{\dy^+}{\dy},
  \quad \EntInnC{\txA}{\rxA}{}{} =\lim_{\dy\downto 1} \EntInnC{\txA}{\rxA}{}{\dy},
\end{align*}
and therefore, $\EntInnC{\txA}{\rxA}{}{} =\lim_{\dy\downto 1} \EntInnC{\txA}{\rxA}{}{\dy}=-\frac{1}{\ratio\tau\beta}\lim_{\dy\downto 1}\frac{\partial}{\partial \dy}\cF(\tau\beta,(\dy-1)\tau\beta).$
\item[6)]
Finally, Theorem 1 in \cite{gaopart1} yields
\comm{
\begin{align*}
    \cIInn{\txA}{\rxA}{} =& -\frac{1}{\ratio\tau\beta}\lim_{\dy\downto 1}\frac{\partial}{\partial \dy}\cF(\tau\beta,\dy\tau\beta)\\
    &+\frac{1}{\ratio\tau\beta}\lim_{\dy\downto 1}\frac{\partial}{\partial \dy}\cF((1+\dy)\tau\beta,0)|_{\dy=1}.
    \numberthis
    \label{eq:IX_Y_tau}
\end{align*}}
\begin{align*}
    \cIInn{\txA}{\rxA}{} =& -\frac{1}{\ratio\tau\beta}\lim_{\dy\downto 1}\frac{\partial}{\partial \dy}\cF(\tau\beta,(\dy-1)\tau\beta)+\frac{1}{\ratio\tau\beta}\frac{\partial}{\partial \dy}\cF(\dy\tau\beta,0)|_{\dy=1}.
    \numberthis
    \label{eq:IX_Y_tau}
\end{align*}
\end{itemize}
By using the expression of $\cF(\betat,\betad)$ in \cite{wen2016bayes}, together with\comm{\eqref{eq:Ent_Y_cond_X}, \eqref{eq:Ent_Y}, and} \eqref{eq:IX_Y_tau}, we have:
\begin{align*}
    \cIInn{\txA}{\rxA}{} 
    = \Enta[]({\snr\qh\qx},\varns+\snr-&\snr\qh\qx)-  \Enta[]({\snr\qh},\varns+\snr-\snr\qh)\\ &+\frac{1}{\ratio}(\IAWGN(\qxtilde,\ptx(\cdot))+\frac{\qx\qxtilde-\qxtilde}{\ln 2}),
    \numberthis
    \label{eq:IXY_value_A}
\\
    \EntInnC{\txA}{\rxA}{ ^+}{} = \Enta[]({\snr\qh}, &\varns+\snr-\snr\qh),
    \numberthis
    \label{eq:HY_X_value_A}
\\
    \EntInnC{\txA}{\rxA}{}{} = \Enta[]({\snr\qh\qx},\varns+\snr-\snr\qh\qx) +&\frac{1}{\ratio}(\IAWGN(\qxtilde,\ptx(\cdot))+\frac{\qx\qxtilde-\qxtilde}{\ln 2}),
    \numberthis
    \label{eq:HY_value_A}
\end{align*}
where $\Enta[](\cdot,\cdot)$ is defined in \eqref{eq:Ent_V_o_def} when $f(\cdot)$ is a $\bit$-bit uniform quantizer and is defined in \eqref{eq:linear_Ent_chi} when $f(\cdot)$ is linear with $f(w)=w$, $\IAWGN(\cdot,\cdot)$ is defined in \eqref{eq:MuI_AWGN_def}, $\qh$ is the solution of  
\comm{
\begin{subequations}
\begin{align}
    \qhtilde=\tau&\beta\snr\cdot\chi(\snr\qh,\varns+\snr-\snr\qh),\\
    \qh&=1-\mse(\qhtilde,\ptg(\cdot)),
\end{align}
\label{eq:qh_qhtilde_solu_A}
\end{subequations}
}
\begin{align}
    \qhtilde=\tau&\beta\snr\cdot\chi(\snr\qh,\varns+\snr-\snr\qh),\qquad
    \qh=1-\mse(\qhtilde,\ptg(\cdot)), \qquad 1-\qh = \msea{\chm},
\label{eq:qh_qhtilde_solu_A&mse_G_vs_qh}
\end{align}
where $\chi(\cdot,\cdot), \mse(\cdot,\cdot)$ are defined in \eqref{eq:chi_value}, \eqref{eq:MSE_AWGN_def}, and $(\qx,\qxtilde)$ are the solution of
\begin{align}
    \qxtilde=\ratio\snr\qh&\cdot\chi({\snr\qh\qx},\varns+\snr-\snr\qh\qx), \qquad \qx=1-\mse(\qxtilde,\ptx(\cdot)).
\label{eq:qx_qxtilde_solu_A}
\end{align}

\subsection{Right sides of \eqref{eq:equi_in_MuI}--\eqref{eq:equi_in_Ent_Y}}
\label{sec:Ent_known_G}
When $\chm$ is known, the entropy $\lim\limits_{\Tlim\to\infty}\frac{1}{\rn}\Ent(\rxvequ|\chm)$ can be computed through 
\begin{equation}
\begin{split}
        \lim\limits_{\Tlim\to\infty}&\frac{1}{\rn}\Ent(\rxvequ|\chm)=-\frac{1}{\ratio\beta}\lim_{\tn\to\infty}\frac{1}{\tn^2}\E\log_2 p(\rxmequ_{\ceil{\beta\tn}}|\chm)
  \\
    &=-\frac{1}{\ratio\beta}\left(-\ratio\beta\Enta[](\snrequ\qxequ,\varnsequ+\snrequ-\snrequ\qxequ)-\beta\IAWGN(\qxtildeequ,\ptx(\cdot))+\frac{\beta(1-\qxequ)\qxtildeequ}{\ln2}\right),
\end{split}
      \label{eq:entropy_known&known_G_block_entropy}
\end{equation}
where $\lim\limits_{\tn\to\infty}\frac{1}{\tn^2}\E\log_2 p(\rxmequ_{\ceil{\beta\tn}}|\chm)$ is available in \cite{wen2016bayes} and $(\qxequ,\qxtildeequ)$ are the solutions of
\begin{align}
    \qxtildeequ=&\ratio\snrequ\cdot\chi({\snrequ\qxequ},\varnsequ+\snrequ-\snrequ\qxequ),\qquad\qxequ=1-\mse(\qxtildeequ,\ptx(\cdot)).
    \label{eq:qx_solu_known_G}
\end{align}

Therefore, \eqref{eq:entropy_known&known_G_block_entropy} yields
\begin{align*}
    &\lim\limits_{\Tlim\to\infty}\frac{1}{\rn}\Ent(\rxvequ|\chm)=\Enta[](\snrequ\qxequ,\varnsequ+\snrequ-\snrequ\qxequ)+\frac{1}{\ratio}\left(\IAWGN(\qxtildeequ,\ptx(\cdot))+\frac{\qxequ\qxtildeequ-\qxtildeequ}{\ln 2}\right).
    \numberthis
    \label{eq:entropy_yt_known_G}
\end{align*}
Since the elements of $\chm$ are \iid $\cC\cN(0,1)$, for any given $\txvequ$, the elements of $\sqrt{\frac{\snrequ}{\tn}}\chm\txvequ$ are \iid $\cC\cN(0,\frac{\snrequ\txvequ^\He\txvequ}{\tn})$. Also, since the elements of $\txvequ$ are \iid with zero mean and unit variance, $\frac{\snrequ\txvequ^\He\txvequ}{\tn}$ converges to $\snrequ$ and the elements of $\sqrt{\frac{\snrequ}{\tn}}\chm\txvequ$ converge to \iid $\cC\cN(0,\snrequ)$ as $\tn\to\infty$.  Therefore, 
\begin{align*}
    &\lim_{\Tlim\to\infty}\frac{1}{\rn}\Ent(\rxvequ|\chm,\txvequ) =\Enta[]({\snrequ},\varnsequ).
    \numberthis
    \label{eq:Ent_Y_cond_X_know_G}
\end{align*}

Then, \eqref{eq:entropy_yt_known_G} and \eqref{eq:Ent_Y_cond_X_know_G} yield
\begin{align*}
   & \lim_{\Tlim\to\infty}\frac{1}{\rn}\MuI(\txvequ;\rxvequ|\chm) = \Enta[](\snrequ\qxequ,\varnsequ+\snrequ-\snrequ\qxequ)-\Enta[]({\snrequ},\varnsequ)+\frac{1}{\ratio}\left(\IAWGN(\qxtildeequ,\ptx(\cdot))+\frac{\qxequ\qxtildeequ-\qxtildeequ}{\ln 2}\right)
    \numberthis
    \label{eq:MuI_know_G}
\end{align*}
where $(\qxequ,\qxtildeequ)$ are the solutions of  \eqref{eq:qx_solu_known_G}. Notice that $\qxtildeequ = \qxtilde$ and $\qxequ = \qx$, and combine with \eqref{eq:SNR_var_equivalent}, we can see that \eqref{eq:entropy_yt_known_G}--\eqref{eq:MuI_know_G} are the same as \eqref{eq:IXY_value_A}--\eqref{eq:HY_value_A}, and thus finish the proof of Theorem~\ref{thm:equivalence_in_Ent}.

\section{$\Ropt$ and $\tauopt$ for Large $\ratio$}
\label{app:large_alpha_rate_and_training}
To prove \eqref{eq:large_alpha_rate_saturation}--\eqref{eq:tauopt_one_bit_output}, we analyze $\bit=\infty$ and $\bit=1$ separately.
\subsection{Large $\ratio$ with $\bit=\infty$}
\label{sec:largea_binf}

$\cIInn{\txA}{\rxA}{}$ can be computed by following the steps in Section \ref{sec:step_by_step_computation}. When $\bit=\infty$,  \eqref{eq:IXY_value_A} yields
\comm{
\begin{align*}
    \ratio\cIInn{\txA}{\rxA}{} &= \frac{1}{\ln 2}\Big(\ratio\ln\left(1+\frac{\snrequ}{\varnsequ}\msea{\tx}\right)-\qxtilde\msea{x}\Big)\\
    &+\IAWGN(\qxtilde,\ptx(\cdot)),
    \numberthis
    \label{eq:IXY_linear_output_temp1}
\end{align*}}
\begin{align*}
    \ratio\cIInn{\txA}{\rxA}{} &= \frac{1}{\ln 2}\Big(\ratio\ln\left(1+\frac{\snrequ}{\varnsequ}\msea{\tx}\right)-\qxtilde\msea{x}\Big)+\IAWGN(\qxtilde,\ptx(\cdot)),
    \numberthis
    \label{eq:IXY_linear_output_temp1}
\end{align*}
where $\ptx(x)$ is the distribution of real/imaginary part of elements of $\txv_t$, $\snrequ, \varnsequ$ are computed from \eqref{eq:SNR_var_equivalent} with $\msea{\chm}$ being the solution of \eqref{eq:qh_qhtilde_solu_A&mse_G_vs_qh}, and $\msea{\tx}, \qxtilde$ are the solution of \eqref{eq:qx_qxtilde_solu_A}, which is 
\comm{
\begin{subequations}
\begin{align}
    \qxtilde&=\frac{\ratio\frac{\snrequ}{\varnsequ}}{1+\frac{\snrequ}{\varnsequ}\msea{x}},\label{eq:qx_qxtilde_solu_linear_fx_a}
    \\
    \msea{x}&=\mse(\qxtilde,\ptx(\cdot)).
    \label{eq:qx_qxtilde_solu_linear_fx_b}
\end{align}
\label{eq:qx_qxtilde_solu_linear_fx}%
\end{subequations}
}
\begin{align*}
    \qxtilde&=\frac{\ratio\frac{\snrequ}{\varnsequ}}{1+\frac{\snrequ}{\varnsequ}\msea{x}},\qquad
    \msea{x}=\mse(\qxtilde,\ptx(\cdot)).
    \numberthis
\label{eq:qx_qxtilde_solu_linear_fx}%
\end{align*}
Then, \eqref{eq:IXY_linear_output_temp1} becomes
\comm{
\begin{align*}
    \ratio\cIInn{\txA}{\rxA}{} &= \frac{\ratio}{\ln 2}\left(\ln\left(1+\frac{\snrequ}{\varnsequ}\msea{x}\right)-\frac{\frac{\snrequ}{\varnsequ}\msea{x}}{1+\frac{\snrequ}{\varnsequ}\msea{x}}\right)\\
    &+\IAWGN(\qxtilde,\ptx(\cdot)).
    \numberthis
    \label{eq:IXY_linear_output}
\end{align*}}
\begin{align*}
    \ratio\cIInn{\txA}{\rxA}{} &= \frac{\ratio}{\ln 2}\left(\ln\left(1+\frac{\snrequ}{\varnsequ}\msea{x}\right)-\frac{\frac{\snrequ}{\varnsequ}\msea{x}}{1+\frac{\snrequ}{\varnsequ}\msea{x}}\right)+\IAWGN(\qxtilde,\ptx(\cdot)).
    \numberthis
    \label{eq:IXY_linear_output}
\end{align*}

The mean-square error of the MMSE estimate is $\mse(\qxtilde,\ptx(\cdot))$ defined in \eqref{eq:MSE_AWGN_def}, which is upper-bounded by the mean-square error of the LMMSE estimate, therefore we get 
\begin{align}
    0\leq \mse(\qxtilde,\ptx(\cdot))\leq \frac{1}{1+\qxtilde},
    \label{eq:MSE_vs_LMMSE}
\\
\msea{x}<\frac{1}{\qxtilde},\quad 0\leq \frac{\snrequ}{\varnsequ}\msea{x}\leq \frac{1}{\ratio-1}.
    \label{eq:bound_mse_x}
\end{align}

Because $\ln(1+w)-\frac{w}{1+w}$ is monotonically increasing in $w$ for $w\geq 0$, \eqref{eq:bound_mse_x} yields $0\leq \ratio\left(\ln\left(1+\frac{\snrequ}{\varnsequ}\msea{x}\right)-\frac{\frac{\snrequ}{\varnsequ}\msea{x}}{1+\frac{\snrequ}{\varnsequ}\msea{x}}\right)\leq \ratio(\frac{1}{\ratio-1}-\frac{1}{\ratio})=\frac{1}{\ratio-1}.$
Therefore,  \eqref{eq:IXY_linear_output} yields $\IAWGN(\qxtilde,\ptx(\cdot))\leq \ratio\cIInn{\txA}{\rxA}{} \leq \frac{1}{\ln 2(\ratio-1)}+ \IAWGN(\qxtilde,\ptx(\cdot)),$
thus
\begin{align}
     \lim_{\ratio\to\infty}\ratio\cIInn{\txA}{\rxA}{}=\lim_{\ratio\to\infty}\IAWGN(\qxtilde,\ptx(\cdot)).
     \label{eq:linear_output_large_ratio_MuI}
\end{align}

As $\ratio\to\infty$, \eqref{eq:qx_qxtilde_solu_linear_fx} and \eqref{eq:bound_mse_x} imply that $\qxtilde\to\infty$, and therefore $\IAWGN(\qxtilde,\ptx(\cdot))\to\Ent(\tx)$ which is the entropy of $\tx$.
For $2^{2\txbit}-$QAM moduation at the transmitter generated by $\txbit$-bit DAC's, we have $\lim_{\ratio\to\infty}\ratio\cIInn{\txA}{\rxA}{}=\Ent(\tx)=2\txbit$
for any finite $\txbit$, and \eqref{eq:opt_rate_per_tx}, \eqref{eq:tau_opt_value} then yield 
\begin{align}
    \lim_{\ratio\to\infty}\Ropt=2\txbit,\quad \lim_{\ratio\to\infty}\tauopt = 0.
    \label{eq:large_alpha_rate_limit_temp}
\end{align}
This shows \eqref{eq:large_alpha_rate_saturation}.

When $\txbit=1$, $\mse(\qxtilde,\ptx(\cdot))$ is upper-bounded by the MSE obtained through a hard decision, or
\begin{align}
    \msea{x}=\mse(\qxtilde,\ptx(\cdot))\leq 4Q(\sqrt{\qxtilde}).
    \label{eq:upp_bound_msex}
\end{align}
For large $\ratio$, \eqref{eq:qx_qxtilde_solu_linear_fx} and \eqref{eq:upp_bound_msex} imply that $\msea{\tx}$ decays exponentially to zero, and therefore
\begin{align*}
    \qxtilde&\approx\ratio{\snrequ}/{\varnsequ},
    \numberthis
    \label{eq:qxtilde_large_alpha_approx}
\\
    \ratio\cIInn{\txA}{\rxA}{} = \frac{\ratio}{2\ln 2}\left(\frac{\snrequ}{\varnsequ}\msea{x}\right)^2&+\IAWGN(\qxtilde,\ptx(\cdot)) + o(\msea{x}^2).
    \numberthis
    \label{eq:IXY_large_alpha_approx}
\end{align*}
Equation \eqref{eq:large_alpha_rate_limit_temp} implies that $\tauopt$ is small when $\ratio$ is large, and therefore, in the following approximations, we only keep the dominant terms in $\tau$. Equations \eqref{eq:qh_qhtilde_solu_A&mse_G_vs_qh} and \eqref{eq:SNR_var_equivalent} yield $\qh\approx\frac{\snr}{1+\snr}\tau\beta$ and $\frac{\snrequ}{\varnsequ}\approx\left(\frac{\snr}{1+\snr}\right)^2\tau\beta$,
and \eqref{eq:qxtilde_large_alpha_approx} then yields $\qxtilde\approx {\snr}^2\tau\beta\ratio/\left({1+\snr}\right)^2$.
It can be shown when $\txbit=1$ that for some $\nu>0$, $\ratio\cIInn{\txA}{\rxA}{}\approx 2-\nu e^{-\frac{\qxtilde}{2}}\sqrt{\qxtilde}$, and that therefore $\tauopt\approx\argmax_{\tau}(1-\tau)(2-\nu e^{-\frac{\nu_1\ratio\tau}{2}}\sqrt{\nu_1\ratio\tau})$ where $\nu_1=(\frac{\snr}{1+\snr})^2\beta$.  Taking the derivative with respect to $\tau$ and setting it equal to zero produces \eqref{eq:tauopt_linear_output}.
\subsection{Large $\ratio$ with $\bit=1$}
We again compute $\cIInn{\txA}{\rxA}{}$ by following the steps in Section \ref{sec:step_by_step_computation}. When $\bit=1$,  \eqref{eq:IXY_value_A} yields
\begin{align*}
    &\ratio\cIInn{\txA}{\rxA}{}=4\ratio\int_{\cR}dz\frac{e^{-\frac{z^2}{2}}}{\sqrt{2\pi}}[ Q\left(\sqrt{{\bar{c}}}z\right)\log_2 Q\left(\sqrt{{\bar{c}}}z\right)\\
    &-Q\left({A}z\right)\log_2 Q\left({A}z\right)]-\frac{\qxtilde\msea{x}}{\ln2}+\IAWGN(\qxtilde,\ptx(\cdot)),
    \numberthis
    \label{eq:one_bit_tx_one_bit_rx_rate_2}
\end{align*}
where
\begin{align*}
    {\bar{c}} = \frac{\snrequ}{\varnsequ},\quad {A}=\sqrt{\frac{{\bar{c}}(1-\msea{x})}{1+{\bar{c}}\msea{x}}},
\end{align*}
$\snrequ,\varnsequ$ are computed from  \eqref{eq:SNR_var_equivalent} which are not functions of $\ratio$, and ($\msea{\tx},\qxtilde$) are the solution of \eqref{eq:qx_qxtilde_solu_A}, which can be expressed as
\begin{align}
    \qxtilde=\int_{\cR}dz&\frac{e^{-\frac{z^2}{2}}}{\sqrt{2\pi}}\frac{\ratio {\bar{c}}}{\pi(1+{\bar{c}}\msea{\tx})}\frac{  e^{- {A}^2 z^2}}{Q({A} z)},
    \quad \msea{\tx}=\mse(\qxtilde,\ptx(\cdot)).
    \label{eq:qx_tilde_one_bit_value}
\end{align}

Since $0<Q(Az)<1$, \eqref{eq:qx_tilde_one_bit_value} yields
\begin{align*}
    &\qxtilde\geq \frac{\ratio {\bar{c}}}{\pi(1+{\bar{c}}\msea{\tx})}\int_{\cR}dz\frac{e^{-\frac{1+2{A}^2}{2}z^2}}{\sqrt{2\pi}}=\frac{\ratio {\bar{c}}\sqrt{1+2{A}^2}}{\pi(1+{\bar{c}}\msea{\tx})}\geq\frac{{\bar{c}}\ratio}{\pi(1+{\bar{c}})}.
    \numberthis
    \label{eq:qxtilde_larger_ratio}
\end{align*}
Similar to \eqref{eq:MSE_vs_LMMSE}, we have $0\leq \mse(\qxtilde,\ptx(\cdot))=\msea{x}\leq \frac{1}{1+\qxtilde}.$
Then, \eqref{eq:qx_tilde_one_bit_value} and \eqref{eq:qxtilde_larger_ratio} yield
\begin{align*}
    0\leq \msea{\tx} <\frac{1}{\qxtilde}\leq \frac{\pi(1+{\bar{c}})}{{\bar{c}} }\cdot \frac{1}{\ratio}.
\end{align*}
Therefore, $\msea{\tx}$ becomes small for large $\ratio$.  A Taylor expansion of \eqref{eq:one_bit_tx_one_bit_rx_rate_2} obtains
\begin{align*}
    &\ratio\cIInn{\txA}{\rxA}{}=\frac{\ratio}{\pi\ln 2}\int_{\cR}dz\frac{e^{-\frac{1+2{A}^2}{2}z^2}}{\sqrt{2\pi}Q({A}z)}\frac{3{A}^2+2{A}-1}{(1+{A}^2)(1+({{A}+{\epsilon}})^2)}{\epsilon}^2+\IAWGN(\qxtilde,\ptx(\cdot)) +O(\ratio\msea{\tx}^2),
\end{align*}
where ${\epsilon} = \frac{(1+{\bar{c}}){\bar{c}}\msea{\tx}}{(\sqrt{{\bar{c}}}+{A})(1+{\bar{c}}\msea{\tx})}$.
Thus, $ \lim_{\ratio\to\infty}\ratio\cIInn{\txA}{\rxA}{}=\lim_{\ratio\to\infty}\IAWGN(\qxtilde,\ptx(\cdot))$.

The remaining steps are similar to $\bit=\infty$ in Appendix \ref{sec:largea_binf} and are omitted.

\bibliographystyle{IEEEtran}
\bibliography{bib/refs.bib}

\end{document}